\documentclass[12pt,preprint]{aastex}
\usepackage{footnote}
\usepackage{graphicx}

\shorttitle{Mass and energy of erupting solar plasma}
\shortauthors{Lee et al.}

\begin{document}

\title{
\bf{Mass and energy of erupting solar plasma observed with the X-Ray Telescope on Hinode}}


\author{Jin-Yi Lee$^1$, John C. Raymond$^2$, Katharine K. Reeves$^2$, Yong-Jae Moon$^{1,3}$, and Kap-Sung~Kim$^{1,3}$}
\affil{$^1$Department of Astronomy and Space Science, Kyung Hee University, Yongin, Gyeonggi 446-701, Republic of Korea \\
$^2$Harvard-Smithsonian Center for Astrophysics, Cambridge, MA 02138, USA \\
$^3$School of Space Research, Kyung Hee University, Yongin, Gyeonggi 446-701, Republic of Korea \\
} 

\begin{abstract}
We investigate seven eruptive plasma observations by {\it{Hinode}}/XRT.
Their corresponding EUV and/or white light CME features are visible in some events. 
Five events are observed in several passbands in X-rays, 
which allows the determination of the eruptive plasma temperature using a filter ratio method. 
We find that the isothermal temperatures vary from 1.6 to 10~MK. 
These temperatures are an average weighted toward higher temperature plasma. 
We determine the mass constraints of eruptive plasmas 
by assuming simplified geometrical structures of the plasma with isothermal plasma temperatures. 
This method provides an upper limit to the masses of the observed eruptive plasmas in X-ray passbands 
since any clumping causes the overestimation of the mass. 
For the other two events, we assume the temperatures are at the maximum temperature of the XRT temperature response function, 
which gives a lower limit of the masses. 
We find that the masses in XRT, $\sim$3$\times$10$^{13}$ $-$ 5$\times$10$^{14}~\rm g$, 
are smaller in their upper limit than total masses obtained by LASCO, $\sim$1$\times$10$^{15}~\rm g$. 
In addition, we estimate the radiative loss, thermal conduction, thermal, and kinetic energies of the eruptive plasma in X-rays.
For four events, we find that the thermal conduction time scales are much shorter than the duration of eruption. 
This result implies that additional heating during the eruption may be required to explain the plasma observations in X-rays for the four events.
\end{abstract}


\section{Introduction}

Coronal mass ejections (CMEs) are among the most powerful solar events. 
The CMEs transfer solar energetic particles to the Earth and cause geomagnetic storms, 
which can produce severe space weather conditions. Previous studies have found that  
the CME plasmas are strongly heated in the low solar corona.
Ultraviolet Coronagraph Spectrometer (UVCS) on board {\it{Solar and Heliospheric Observatory}} (SOHO) observations 
 have shown that the heating energy of  CME plasma is comparable to or even larger than its kinetic energy 
 \citep{akmal2001, ciaravella2001, lee2009, landi2010, murphy2011}. 
Ion charge composition from in situ measurements by the {\it{Advanced Composition Explorer}} (ACE) have shown that 
strong and rapid heating are required around 2~R$_\sun$ by ionization state models \citep{gruesbeck2011, rakowski2011,lynch2011}. 
Recent studies using the {\it{Solar Dynamic Observatory}}/Atmospheric Imaging Assembly (AIA) observations have shown that 
CME core regions are strongly heated, and their temperatures can be higher than 10 MK \citep{reeves2011, cheng2012, hannah2013, tripathi2013}. 

The association between hot X-ray plasma ejections and CMEs has been studied 
since \citet{harrison1985} who found X-ray plasma ejections to be a signature of CME onset using observations by the 
Coronagraph/Polarimeter \citep{macqueen1980} and the hard X-ray imaging spectrometer \citep{vanbeek1980} 
on board the Solar Maximum Mission. 
\citet{shibata1995} suggested that hot plasma ejections, observed by the soft X-ray telescope (SXT; \citealp{tsuneta1991}) on board {\it{Yohkoh}} \citep{ogawara1991},  
might be miniature versions of the much larger scale CME events. The temperatures and masses of ejected plasmas 
have been estimated to be higher than 10 MK  and about 10$^{13}\sim$10$^{14}$~g, respectively \citep{tsuneta1997, ohyama1997, ohyama1998, tomczak2004}.

\citet{nitta1999} made the first systematic comparison between flare-associated plasma ejections in SXT and CMEs observed by the 
Large Angle and Spectrometric Coronagraph Experiment (LASCO) on board {\it{SOHO}}. 
They proposed that the occurrence of a hot plasma ejection depends on the presence of open field lines created by a preceding CME. 
A statistical study of the relationship between flare-associated X-ray plasma ejections and CMEs shows that the events with X-ray plasma ejections 
are more correlated with CMEs \citep{kim2005a}. 
Recently, a catalogue of X-ray plasma ejections observed by SXT has been developed by \citet{tomczak2012}.
They suggests that the erupting loop-like structure characteristic of X-ray plasma ejections is a promising candidate to be a high temperature precursor of CMEs.
This idea is consistent with an earlier result that the X-ray plasma ejections with both loop and jet-type structures show a high association with CMEs \citep{kim2004}.
A spectroscopic study of hot plasma associated with a CME using the EUV Imaging Spectrometer (EIS) and the X-ray Telescope (XRT) on board {\it{Hinode}} \citep{landi2013} shows that 
the temperature of X-ray plasma ejection is about 8$-$12.5~MK with the absence of the hot spectral lines in EIS, which might come from the limit 
of sensitivity of EIS for the high temperatures.

Several studies have investigated the physical properties of erupting hot X-ray plasmas associated 
with CMEs \citep{tomczak2004, kim2005b, landi2010}.
\citet{tomczak2004} investigated the energy balance of an X-ray plasma ejection associated both an X-class flare and a CME. 
He suggests that the mass of the X-ray ejection event, 1.2$-$1.5$\times$10$^{15}$~g, can be 
responsible for an additional acceleration of the CME. 
A study of kinematic properties shows that the majority of X-ray plasma ejections are not likely to be the X-ray 
counterpart of CMEs but outflows generated by magnetic reconnections with no statistical correlation between the 
speed of X-ray plasma ejections and the corresponding CME speeds \citep{kim2005b}.

In this analysis we investigate energies of eruptive plasmas by estimating the temperatures and masses using the {\it{Hinode}}/XRT observations. 
We find the temperatures using the filter ratio method \citep{narukage2011, dudik2011, kariyappa2011,reeves2012}, 
because there are not enough filters available for a more sophisticated differential emission measure (DEM) analysis for our observed events.
We determine the mass constraints of eruptive plasma by 
calculating the emission measures and assuming simplified geometrical structures to determine the densities.
These assumptions provide the upper limit for the masses of the observed eruptive plasmas. 
The masses of CMEs have been studied previously using the {\it{SOHO}}/LASCO observations \citep[e.g.][]{vour2010}. 
We compare the estimated masses from X-ray observations with those calculated from the coronagraph measurements.
Finally, we investigate the radiative loss, thermal conduction, thermal, and kinetic energies of eruptive X-ray plasma.

In Section 2, we describe the observations used in this analysis. In Section 3, we explain the analysis method. 
In Section 4, we discuss the estimated temperatures, masses, and energies of erupting X-ray plasmas. 
In Section 5, we present our conclusions.

\section{Observations}
\label{sec:obs}

The XRT \citep{golub2007, kano2008} on board {\it{Hinode}} \citep{kosugi2007} observes 
the solar corona at high ($\sim$\,1$''$ pixels) spatial resolution in multiple bandpasses 
with a wide field of view of 34 $\times$ 34 arcmin and a 2048$\times$2048 CCD camera.
We investigate 7 examples of eruptive X-ray plasma observed by XRT (Figure~\ref{fig:xrt} and Table~\ref{tb:xrt}).
Four events are observed at the full resolution with $\sim$\,1$''$ pixels with a field of view (FOV) of $\sim$\,512$''\times$512$''$.  
Two events are observed with 4 $\times$ 4 pixel binning. In this mode, XRT can observe the full Sun in 512$\times$512 pixels with $\sim$\,4 arcsec/pixel resolution. 
One event is observed with a 2$\times$2 pixel binning with a FOV of $\sim$\,1024$\times$1024$''$. 
Five events among the seven events are observed in a few passbands, 
which allows us to determine the temperature of the erupting plasma using the filter ratio method.
We show the observational time cadences in Table~\ref{tb:xrt}. 

Corresponding EUV and/or white light CME features are visible in some events.
Several events are observed in EUV bandpasses by the Extreme ultraviolet Imaging Telescope (EIT; \citealp{dela1995}) on board {\it{SOHO}} 
and the Sun Earth Connection Coronal and Heliospheric Investigation (SECCHI) 
Extreme UltraViolet Imager (EUVI; \citealp{wuelser2004, howard2008}) on board {\it{Solar TErrestrial RElations Observatory}} (STEREO; \citealp{kaiser2008}).
 We summarize the EUV observations for each event in Table~\ref{tb:euv}. 
The EIT and EUVI observations show associated prominence eruptions in absorption or emission for several events with the eruptive plasmas in X-rays. 
Five events are observed in 195~\AA\ by {\it{SOHO}}/EIT with a 12 minute cadence. 
All seven events are observed by {\it{STEREO}}/EUVI with various time cadences from a few minutes to 20 minutes for most of observations. 
Two events are associated with B class flares by the {\it{Geostationary Operational Environmental Satellite}} (GOES). 

Four events are associated with CMEs observed by LASCO \citep{brue1995} on board {\it{SOHO}}, 
while we do not find the associated CMEs for other three events. 
One event on 2008 December 30 among the four events is associated with a very poor CME. 
The EUVI observations on 2009 January 11 show brightening in 171~\AA, 195~\AA, and 284~\AA\ and, 
especially, a brightening along a filament in 304~\AA. 
Observations near the East limb on 2010 January 22 show that the erupting material moves to the southeast along a larger loop in the EIT observation, 
subsequently causing another eruption associated with a CME. 
A large filament is placed on the eruption region on 2010 January 30. 
The 304~\AA\ observations of the event show that some materials move along the large filament.

\section{Analysis}

We estimate the temperature and mass of erupting plasma observed in X-rays,
and compare the mass constraints with the masses 
obtained from white light coronagraph observations by the LASCO.
In addition, we evaluate the radiative loss, thermal conduction, thermal, and kinetic energies of eruptive plasmas observed in the X-rays.

\subsection{Temperature estimation of eruptive X-ray plasma}
\label{sec:temp}

We use the filter ratio method \citep{narukage2011, narukage2014} to find the temperatures of erupting plasma 
for 5 events observed 
in a few passbands by the XRT (Table~\ref{tb:temp}). 
Filter ratio methods using the XRT observations have been used previously to estimate 
the temperatures of bright points \citep{kariyappa2011}, the coronal emission for an active region \citep{dudik2011}, 
and a coronal cavity \citep{reeves2012}.   
In this analysis, the temperatures of the erupting plasma are estimated using relatively thinner filter observations
since the thicker filter observations do not have enough signal for the temperature estimation. 
We exclude the pixels that show contamination spots
 ,which are spot-like patterns over the CCD that accumulated after the second bakeout on 3 September 2007 (see details \citealp{narukage2011}). 
The observed data are calibrated by a procedure (xrt\_prep.pro) in {\it{SolarSoft}} \citep{kobelski2014}. 
We use a procedure (xrt\_teem.pro) in SolarSoft to estimate the temperatures by a filter ratio method \citep{narukage2011, narukage2014}.

Figure~\ref{fig:temp} shows the regions used for the temperature estimation.
 We use the black colored pixels in the fourth panels of Figure~\ref{fig:temp} for the estimation.
 In addition, we estimate the temperatures using averaged intensities for the regions which are included by contours on both the second and third panels of Figure~\ref{fig:temp}.
We subtract a pre-event image for each passband observation before we apply the filter ratio method.
The images at the first column show the eruptive plasmas before the pre-event subtractions. 
The regions of eruptive plasma are shown in the second and third columns for two XRT passbands via  enclosed solid lines on the pre-event subtracted images. 
The time differences of the observations in two passbands are within 1 minute for most of the events. 
We use thresholds for the temperature error and photon noise \citep{narukage2011} of less than 50\% and 20\%, respectively.
The images at the last column show the pixels within the thresholds by a black color on the XRT observations.
We estimate the signal to noise ratio (S/N) by using the procedure xrt\_cvfact.pro in {\it{SolarSoft}}.
We present the averaged S/N values of the selected pixels for the temperature estimation in Table~\ref{tb:temp}.
The corresponding temperatures to the averages of the observed data ratios are shown in Figure~\ref{fig:ratio}.
 The slight difference in the model filter ratios on 2007 May 23 compared with the ratios on 2009 January 22 and 30 
come from small changes of the temperature response due different levels of contamination on the CCD as a function of time. 
The Al/mesh filter observations have been more affected by the contamination \citep{narukage2011}.

In case of two events on 2008 April 9 and 2009 December 13 observed by a single filter, 
we assume the temperature of the eruptive plasma at the maximum of the temperature response function for that filter (See Section~\ref{sec:mass}).

\subsection{Mass estimation of eruptive X-ray plasma}
\label{sec:mass}

We determine the mass constraints of eruptive plasma 
by assuming simplified geometrical structures. 
For most of events in this analysis, the eruptive plasmas expand with a loop-like structure.
We assume a simple cylindrical structure for the loop-like eruptive plasmas.
The widths and lengths of the cylinder structures are shown in Figure~\ref{fig:temp} and also in Figure~\ref{fig:mass}.
The width of the cylinder structure is assumed to equal the line of sight depth. 
With these assumptions, the volume of the eruptive plasma of each event 
can be estimated as a cylinder.

For the event on 2008 April 8 in Figure~\ref{fig:mass}, we select two regions (A and B) of the eruptive plasmas because the region `B' appears to consist of  
ambient corona plasma surrounding region `A'. 
The region `A' of erupting plasma shown in the second column in Figure~\ref{fig:mass} is taken as a blob structure.
In this case, we take the width at each row (w$_{\rm i}$, see the second column in Figure~\ref{fig:mass}), assume a short cylinder at that position, 
and sum the masses of these short cylinders for the total mass of the region `A'.
The region `B' shown in the third column is assumed to be a cylinder structure like the other events.

The above assumptions of the structure provide the upper limit to the masses of the observed eruptive plasmas 
since any clumping will cause an overestimate of the mass.
Once we define the structure of the eruptive plasma for each event, 
we find an emission measure (EM) of the erupting plasma 
which can be used to estimate the densities of the plasmas. 
The observed data number (DN) is given by 

\begin{equation}
DN = R_t (T) \times \rm{EM}, 
\end{equation}

\noindent where DN is normalized number of DN sec$^{-1}$pix$^{-1}$ and R$_{t}$(T) ([DN cm$^5$ sec$^{-1}$pix$^{-1}$]) is the temperature response function
from procedures (make\_xrt\_wave\_resp.pro and make\_xrt\_temp\_resp.pro) in {\it{SolarSoft}}.
The EM [cm$^{-5}$] is defined as 

\begin{equation}
EM = < n_e^2 > dl, 
\end{equation}

\noindent where {\it n$_e$} and {\it dl} are the electron density and line of sight depth of the observed plasma, respectively.
Therefore, we can derive the electron density of the eruptive plasma with the assumption of the line of sight depth as explained above.

For the events in Figure~\ref{fig:temp}, we use the temperature response at the estimated temperatures given by the filter ratio method.
For the two events observed with a single passband, we assume the maximum temperature response for that filter, and the corresponding temperatures 
are assumed as the temperatures of the eruptive plasmas for these events.
Under this assumption, the estimated density gives the lower limit to the mass.
We show the temperature response functions with the estimated (five events using the filter ratio method) or 
assumed (two events with a single filter observation) temperatures in Figure~\ref{fig:resp}.

Finally, we estimate the masses of the eruptive plasmas in X-rays using the averaged density of the selected region for each event. 
For this we assume a plasma with a 10$\%$ of Helium content with a mass of 1.974 $\times$ 10$^{-24}$~g per ion. 
Table~\ref{tb:mass} shows the geometrical parameters, density, and mass for each event. 
We show the S/N values for the regions used for the mass estimation in Table~\ref{tb:mass}. 
These are estimated by the same procedure as the S/N values in the Table~\ref{tb:temp}. 

\subsection{Energies of eruptive X-ray plasma}
\label{sec:energies}

We estimate the radiative loss, thermal conduction, thermal energy, and kinetic energy of the eruptive plasma 
using the physical quantities in Table~\ref{tb:mass}. 
The radiative loss (L$_r$) is given by

\begin{equation}
\rm{L_r} ~[erg] = EM \times P(T) \times \bigtriangleup{t},
\end{equation}
 
\noindent where EM is in unit of cm$^{-3}$ and P(T) is plasma radiative loss function. 
We use  analytical fits, P(T)=10$^{-21.94}$ for temperatures lower than 10$^{6.3}$ K and P(T)=10$^{-17.73}$~T$^{-2/3}$ 
for temperatures higher than or equal to 10$^{6.3}$K, for the radiative loss function given by \citet{rosner1978}.  
The quantity $\bigtriangleup$t is the duration of eruptive plasma (column four in Table~\ref{tb:energy}).
We define the beginning of the duration right after the eruption starts. 
In the case of the event on 2009 December 13, the eruption is first observed at 09:08~UT, but there is a data gap between 08:40~UT and 09:08~UT. 
Therefore, the duration could be longer. The field of view of the event on 2010 January 30 is not enough to follow the eruption. 
It appears that the bright eruptive plasma could persist longer than the duration 
since the eruptive plasma moves out of the FOV in the observations later the duration in Table~\ref{tb:energy}. 
In addition, most of eruptions start from pre-existing bright structures in the XRT observations. 
Thus, the duration of the eruptive plasma could be longer than the time used in this analysis.

The radiative loss time scale \citep[e.g.][]{golub2009} is estimated by 

\begin{equation}
 \tau_{rad} = \frac{3~(n_e+n)~k_B~T}{2~n_e~n~P(T)}, 
\end{equation}

\noindent where n is the number density assuming 0.8 times of the electron number density, k$_B$ is the Boltzmann constant, and T is the temperature. 

We assume that the the thermal conduction (F$_c$) is mainly along the magnetic field lines and estimate the thermal conduction using the conductivity flux defined as   

\begin{equation}
F_c ~ \rm{[erg]} = \kappa~T_{max}^{5/2}~\frac{T_{max}}{{\it{l}}/2}~A \times \bigtriangleup{t},
\end{equation}

\noindent where $\kappa$ is 1.8$\times$10$^{-5}$/$\rm{ln}$$\Lambda$($\approx$ 7.0$\times$10$^{-7}$~erg~cm$^{-1}$~K$^{-7/2}$~s$^{-1}$), 
T$_{max}$ is the temperature at the loop top, $l$ is the length of the loop, and A is the cross-sectional area.   
Here, the \rm{ln}$\Lambda$ is the Coulomb logarithm. We apply the half lengths and depths of the cylinder structures to estimate the thermal conduction.  

The thermal conduction time scale is estimated by 

\begin{equation}
\tau_{cond} = \frac{3~(n_e + n)~k_B~(l/2)^2}{\kappa~{T_{\rm{max}}}^{5/2}}.
\end{equation}

\noindent We use the estimated temperatures in the Section~\ref{sec:temp} for the T$_{\rm{max}}$. 
However, the estimated temperature is during the eruption of the plasma.  
The temperature of the eruptive plasma may decrease due to its expansion. 
Therefore, the estimated thermal conduction time scale can be larger than our estimation.
In the case of the events on 2009 January 11 and 2010 January 22, the length of the cylinder structure of the eruptive plasma in Figure~\ref{fig:temp} 
is the half of the length of the loop. Therefore, the length in Table~\ref{tb:mass} is used for the {\it{l}}/2 in equations~(5) and~(6). 

We estimate the thermal energy (=3(n + n$_e$)k$_B$T $\times$ Volume) and kinetic energy (=1/2mv$^2$) with the mass, m, in Table~\ref{tb:mass}. 
The volume is calculated using the geometrical parameters in Table~\ref{tb:mass}.
The projected speeds of the erupting plasma are estimated using several sequence images during the times (column eight) in Table~\ref{tb:energy}. 
Thus, the kinetic energy could be larger than these estimates. 

In addition, we estimate the potential energy, which is required to lift the erupting plasma against solar gravity, given by 

\begin{equation}
U = GM_\sun m(\frac{1}{R_\sun} - \frac{1}{r}),
\end{equation}

\noindent  where G is the gravitational constant, M$_\sun$ is the solar mass, and R$_\sun$ is the solar radius. 
We define the final heights (r, column eleven in Table~\ref{tb:energy}) of the eruptive plasma from the observations 
just before the plasmas are blown out by losing their structures in the images, except for three events (marked by $^\flat$ in Table~\ref{tb:energy}). 
The FOVs of the XRT observations for the three events are not enough to see the final heights of the eruptive plasmas. 
For these events, we define the final heights as the highest height of eruptive plasma that can be measured in the limited FOV. 
Thus, the final heights could be higher than the measured values.

\section{Results and discussion}
\label{sec:results}

We estimate the temperature, mass, and energies of eruptive plasma for seven events observed in {\it{Hinode}}/XRT.
The temperatures estimated by the filter ratio of the plasma vary from 1.6 to 10~MK in Table~\ref{tb:temp}. 
 Using the black pixels in the fourth column of Figure 2, we find that the event on 2007 May 23 shows a high temperature of $\sim$10~MK using the filter ratio method.
This measurement confirms the results of temperature analysis studies of CME plasmas, which show that 
the temperatures of the CMEs observed in EUV can include high temperature plasma over 10~MK \citep{reeves2011, cheng2012, hannah2013, tripathi2013}. 
The temperature of erupting plasma on 2009 January 22 is $\sim$6.3~MK 
while the temperatures for other three events on 2008 December 30, 2009 January 11, and 2010 January 30 are relatively low, from 1.6 to 2.5~MK.
 Using the areas enclosed in the contours shown in the second and third panels of Figure 2, 
we find temperatures of 16, 2.2, 2.2, 7.9, and 2.5~MK for the events on 2007 May 23, 2008 December 30, 2009 January 11, 2010 January 22, and 2010 January 30, respectively.
The estimated temperatures are higher than or similar to the temperatures estimated by using the black pixels as explained in section 3.1.
Lower DNs in faint regions of erupting plasma within the contours might be removed by the background subtraction. 
This effect is more likely for Al/mesh filter observations, which are relatively more sensitive for the lower temperature plasma, meaning that 
the contribution of lower intensity pixels could be disproportionately prevalent in the Al/mesh filter.  
Thus the average Al/mesh intensity could be lower than it should be, which would increase the filter ratio, thereby increasing the measured temperature.

Two events on 2007 May 23 and 2009 December 13 are associated with B-class flares. 
The temperature of erupting plasma on 2007 May 23 is higher than other events. 
The event on 2009 December 13 is observed with a single filter of Al-med, which is sensitive for high temperature ($> \sim$5~MK).
Therefore, this observation may indicate the high temperature of the erupting plasma. 
In addition, the source location of the event on 2008 April 9 is behind of the limb. 
This event has been studied with regard to supra-arcade downflows observed by XRT 
that have been interpreted as evidence for reconnection in a current sheet trailing a flux rope eruption \citep{savage2010}. 
Therefore, there is a possibility of a flare behind the limb. We can not see an evidence of the flare in the STEREO observations 
since the active region is seen at the west limb by the STEREO at heliocentric longitude 24$^\circ$. 
One event on 2010 January 22 shows a relatively high temperature of $\sim$6.4~MK without an associated flare.
The results of our study suggest that the events of erupting plasma in X-rays can be seen without any flare. 
However, the higher temperature plasma events are possibly more associated with flares 
as found in a previous study using observations by SXT on board {\it Yohkoh} by \citet{tomczak2012}.

The masses estimated by using a simplified geometrical structure are from $\sim$3$\times$10$^{13}$ $-$ $\sim$5$\times$10$^{14}$~g. 
The averaged densities of eruptive plasma are in the range between $\sim$7$\times$10$^7$ and $\sim$2$\times$10$^9~\rm cm^{-3}$. 
These results are smaller than earlier studies using {\it{Yohkoh}} observations \citep{tsuneta1997, ohyama1997, ohyama1998, ohyama2008, tomczak2007}
probably because their X-ray plasma ejections are associated with mostly M-class flares. 
Several of our events are cooler, but the hotter events have temperatures similar to those listed in \citet{tomczak2007}.
The event on 2008 April 9 has been intensively studied on its energetics and physical properties of current sheet and flux rope  
\citep{savage2010, landi2010, patsourakos2011,  thompson2012, landi2013}.
The density, 8.8$\times$10$^7$ cm$^{-3}$, in this analysis is similar with the result, $\sim$10$^8$ cm$^{-3}$, of \citet{landi2010}. 
The temperature, log T = 6.9 $-$ 7.1 K, from the analysis using the EIS and XRT \citep{landi2013} is 
similar with our assumption at the maximum temperature response in this analysis.
The masses of the eruptive plasma in lower corona are smaller than those calculated from coronagraph observations by LASCO, as shown in Table~\ref{tb:euv}. 
The coronagraph observations represent the total mass of CME materials measured through electron scattering, 
while the mass in X-rays represents only the hot plasma. There might also be mass that falls down to the Sun.
In addition, the CMEs accumulate mass including cooler plasma as they expand through the corona, 
so observations at 2.3 R$_\sun$ and above should give larger masses.

Three events among the seven events have no corresponding CME observations by LASCO.
The event on 2010 January 22 shows a failed eruption in the EIT and EUVI observations. 
The observations in EUV on 2010 January 30 show a large filament near the region where the eruption is observed. 
The 195~\AA\ EUVI observations appear to show eruptions on 2010 January 30. 
However, the observations show that some of the erupting plasma falls back to the Sun.
The event on 2009 January 11 also shows a brightening along a filament in 304~\AA\ EUVI observations. 
Thus, this event also has a possibility that some plasmas might fall back to the Sun along the filament.

Two events on 2007 May 23 and 2009 December 13 are associated with CMEs in the LASCO observations.
The event on 2008 December 30 is very slow ($\sim$5 km/s) in the low corona. 
The event is possibly associated with a very faint narrow CME appeared in C2 FOV at 21:30~UT about 3 hours later than the event observation time in XRT. 
Only about half of the eruptions in X-rays are associated with the CMEs in the LASCO
observations, either because the erupting material falls back to the Sun or because
the ejected mass is too small to be seen given the sensitivity and cadence of LASCO.

Two events have corresponding EUV observations by {\it the Transition Region and Coronal Explorer} (TRACE; \citealp{handy1999}) available for comparison.
Figure~\ref{fig:trace} shows the co-alignment between the TRACE and XRT observations.
The event on 2007 May 23 shows the eruptive plasma as an absorption from $\sim$07:12~UT to $\sim$07:42~UT  in 171~\AA\ and 195~\AA\ by {\it{STEREO}}/EUVI. 
The 304~\AA\ EUVI observations show the eruption in emission at $\sim$07:22~UT, but later, it appears in absorption in the same 304~\AA\ band. 
The eruption in X-rays starts at $\sim$07:02~UT and then all the plasma erupts before $\sim$07:20~UT.
This indicates that the eruptive hot plasma in X-ray cools down, so then it appears as absorption in 195~\AA\, then in 304~\AA\ as emission and finally as absorption in 304~\AA. 
The co-alignment image on the left bottom in Figure~\ref{fig:trace} demonstrates that 
the erupting plasma observed by both TRACE and XRT is spatially coincident. 
The erupting loop on the TRACE observations contains a filament which is seen as an absorption feature on the {\it STEREO}/EUVI observations.
The EUVI 171 \AA\ observations also show a loop eruption in emission.
Thus, it is also possible that a filament erupts with hot plasma which can be seen in emission.

On the other hand, the event on 2008 Apr. 9 shows that a prominence erupts in absorption around 09:05~UT in 171~\AA, 195~\AA, and 284~\AA\ observations 
and then it appears as emission in the passbands, while it erupts in emission in 304~\AA\ and remains in emission until it moves out of the observational FOV.
Its eruption in X-ray starts around 09:16~UT from the behind of the limb. 
Therefore, the eruption might start earlier than that time.
These observations indicate that the hot and cool plasmas erupt simultaneously and the cool plasmas seen in absorption are heated during their eruptions.
The co-alignment image on the right bottom in Figure~\ref{fig:trace} shows that 
the plasmas seen in the TRACE and XRT observations are not spatially coincident. 
Landi et al. interpret this observation as the XRT emission enveloping the tail of the erupted prominence.

Table~\ref{tb:energy} shows the radiative loss, thermal conduction, thermal, and kinetic energies. 
The radiative loss energies are much smaller than the thermal conduction energies, 
except for one event on 2010 January 30. 
In this event, the radiative loss is larger than thermal conductive flux by a factor of 3-4.  
We estimate the temperatures in this events to be the relatively low values of 1.6 MK 
and the radiative loss time scale is smaller than the thermal conduction time scale. 
Additionally, the event has much smaller thermal conduction flux than its thermal energies.  
On the other hand, for two events on 2007 May 23 and 2009 December 13, we estimate high temperatures of 10 MK.   
These events have relatively large thermal conductive flux compared to their thermal energies, 
and their conductive cooling times are very short compared to the radiative cooling times. 
Therefore, the plasma of the two events with the high temperatures of 10 MK is dominated by thermal conduction, 
while the plasma of the two relatively low temperature events is dominated by radiative losses. 
Three other events on 2008 April 9, 2009 January 11, and 2010 January 22 show similar conduction and thermal energies, 
while one event on 2008 December 30 shows a relatively larger thermal conduction than its thermal energy.
These events are also dominated by the conductive cooling process, according to the calculated cooling times.

The thermal energies are much larger than the kinetic energies for all events from X-ray observations in low corona.
The kinetic energies from the observations in X-rays are much smaller than those of typical CMEs from the LASCO observations 
because of their low speeds and masses.
Therefore, the eruptive plasma in the lower corona might be accelerated during its expansion to 
the several solar radii when it is observed by coronagraphs. 
The potential energies are higher than the kinetic energies 
because of the very low speeds and masses in low corona. 
On the other hand, the potential
energies are smaller than the thermal energies because the temperatures
are high and the gas is only lifted 0.1$-$0.3 R$_\sun$ above the solar surface.
\citet{emslie2012} studied energetics of large solar eruptive events.
In their study, the kinetic energies are much bigger than the potential energies ($\sim$10$^{30} - $10$^{31}$~erg), which are estimated from the LASCO observations. 
The kinetic energies of their events, $\sim$10$^{31} - $10$^{33}$~erg, are larger than those of our three events, $\sim$10$^{30}$~erg, 
taken from the LASCO CME catalogue\footnote{\label{catalogue}http://cdaw.gsfc.nasa.gov/CME$\_$list/}.
The events studied in \citet{emslie2012} are mostly associated with X-class flares, 
and the potential and kinetic energies are estimated at larger heights. 
Thus, the kinetic energies would be bigger than the potential energies if the speed exceeds the escape speed.

In addition, we investigate the energy evolution of two of the five events (2007 May 23 and 2008 December 30) 
observed with multiple passbands.
Three other events are not available for this analysis 
due to the lower cadence of the second filter observations and a limitation of the FOVs. 
Especially, the event on 2008 December 30 is very slow with a velocity of about 3$\sim$7~km/s for about 4 hours. 
This slow expansion makes it possible to see the eruption for a longer time in the XRT field of view.
We apply the velocities in Table~\ref{tb:energy} for both events since the velocities are similar in the time sequences. 

Figure~\ref{fig:energy} shows the energy changes for two events. 
The event on 2007 May 23 is observed for only three times. 
The erupting plasma seems to heat up (or at least not cool down) even as the structure is expanding. 
The thermal conduction, thermal, and kinetic energies increase as the structure and temperature become larger.
The event on 2008 December 30 shows a brightening at the beginning of its eruption. 
The temperature of the plasma is higher, at about 9~MK at 14:41 UT. 
Then, the erupting plasma cools down, but the temperature stays larger than 2~MK even after its slow decrease between 15:06 and 16:52 UT.
The kinetic and thermal energies increase with time as the volume and mass get larger.
For both events, we confirm that the thermal conduction time scales remain much shorter than the radiative loss time scales during expansion.

With above results, we confirm that the radiative loss and thermal conduction could cool the erupting hot plasma. 
Thermal conduction does even out the temperature along the
loop.  The reason that it is considered a cooling process is that
it transports heat from the X-ray emitting gas to temperatures
of a few times 10$^5$ K.  The radiative cooling coefficient in this temperature range is
much higher, and the density is also higher, so the transition region
radiates the energy away very efficiently.
Especially, for 4 events on 2007 May 23, 2008 April 9, 2008 December 30, and 2009 December 13, 
the thermal conduction timescales are much shorter than the duration of eruption (column four in Table~\ref{tb:energy}).
This result implies that additional heating energy for these four events is necessary 
to explain the persistence of the high temperature plasma in X-ray observations during the eruption. 

\section{Summary and Conclusion}

We investigate the temperatures, masses, and energies of seven eruptive plasma observed by {\it{Hinode}}/XRT. 
We estimate the temperatures of X-ray plasma by a filter ratio method for five events observed by several passbands. 
For other two events observed by a single passband, we assume the temperatures at the maximum of the temperature response
function for that filter. We find that the estimated isothermal temperatures vary from 1.6~MK to 10~MK. 

About half of the erupting plasmas have no corresponding CME observed by the LASCO due to failed eruption  
or the sensitivity and cadence of the LASCO. 
We determine the mass constraints of eruptive plasmas in X-rays assuming simplified cylinder structures, which represent an upper limit to the mass. 
The estimated masses are 3$\times$10$^{13} - $5$\times$10$^{14}$~g. 
The masses in X-rays are smaller than 
the masses in coronagraph observations.
The estimated mass in X-rays represents only the hot plasma while the mass from LASCO 
is total CME mass. In addition, the CMEs accumulate mass including cooler plasma as they expand through the corona.

The radiative loss energies are much smaller than the thermal conduction energies in six events except for 2010 January 30. 
Therefore, the plasmas for most of events cool down by the thermal conduction, while the plasmas for only one event cools down due to radiative loss.
The thermal conduction time scales for four events are much shorter than the duration of eruption. 
This result indicates that additional heating for these four events is required to explain the hot plasmas observed by {\it{Hinode}}/XRT.
To investigate how much heating energy will be required to explain the hot plasma in X-rays 
and the kinetic energies of the CMEs observed by coronagraphs, 
we plan to analyze the time series observations in X-rays with high cadence {\it{SDO}}/AIA images.

\acknowledgments

This work was supported by NASA grants NNM07AA02C and NNX09AB17G and NSF SHINE grant AGS-1156076 to the
Smithsonian Astrophysical Observatory, Basic Science Research
program (NRF-2013R1A1A2058409, NRF-2013R1A1A2012763) and the BK21 plus program through the National Research
Foundation (NRF) funded by the Ministry of Education of Korea, 
NRF of Korea Grant funded by the Korean Government (NRF-2013M1A3A3A02042232), 
and the Korea Meteorological Administration/National Meteorological Satellite
Center. 
``Hinode is a Japanese mission developed and launched by ISAS/JAXA, with NAOJ as
domestic partner and NASA and STFC (UK) as international partners. It is
operated by these agencies in co-operation with ESA and the NSC (Norway)."
The CME catalog is generated and maintained at the CDAW Data Center by 
NASA and The Catholic University of America in cooperation with the Naval Research Laboratory. 
SOHO is a project of international cooperation between ESA and NASA. 

\bibliographystyle{apj} 
\bibliography{ms1}

\clearpage

\begin{deluxetable}{lllccl} 
\tabletypesize{\footnotesize}
\tablecaption{XRT observations} 
\tablewidth{-100pt}
\tablehead{ \colhead{Date} & \colhead{Location} &  \colhead{Filter} & \colhead{Bins} & \colhead{Cadence$^a$}  &  \colhead{GOES X-ray$^b$}  }

\startdata
2007/05/23  & (~770\arcsec, ~-10\arcsec) & Ti\_poly, Al\_mesh & 1 & 20 sec  & B5.3 (S10W51)  \\
2008/04/09  & (~890\arcsec, -350\arcsec) & Al\_poly & 1 & 1 min  & None \\
2008/12/30  & (-600\arcsec, -400\arcsec)  & Be\_thin, C\_poly, Be\_med & 4  & 8 min & None \\
2009/01/11  & (-100\arcsec, ~400\arcsec) & Be\_thin, C\_poly, Be\_med & 4  & 8 min  & None\\
2009/12/13  & (-400\arcsec, ~350\arcsec)  & Al\_med & 2 & 6 min &  B1.4 (N22E26)\\
2010/01/22  & (-720\arcsec, -420\arcsec)  & Ti\_poly, Al\_mesh, Al\_thick & 1 & 1 min & None\\
2010/01/30  & (~580\arcsec, -420\arcsec) & Ti\_poly, Al\_mesh & 1 & 1 min  & None
\enddata
\tablecomments{\\
$^a$Observational time cadence for the first filter in the third column.\\
$^b$GOES X-ray flare from http://www.lmsal.com/solarsoft/latest\_events/. }
\label{tb:xrt}
\end{deluxetable}

\begin{deluxetable}{lcllll} 
\tabletypesize{\footnotesize}
\tablecaption{Temperature in X-ray passbands} 
\tablewidth{-100pt}
\tablehead{ \colhead{Date} & \colhead{Time (UT)} & \colhead{S/N} &
\colhead{Filters} & \colhead{Temp.}}

\startdata
2007/05/23 & 07:10 & 5.5 & Ti\_poly/Al\_mesh & ~10 MK \\
2008/04/09 & 09:52 &        &  Al\_poly  &  ~~8 MK              \\
2008/12/30 & 18:30 & 6.1 & Be\_thin/C\_poly & 2.5 MK \\
2009/01/11 & 23:21 & 3.7 & Be\_thin/C\_poly & 2.5 MK \\
2009/12/13 & 09:13 &        &   Al\_med &  ~10 MK            \\
2010/01/22 & 14:05 & 2.7 & Ti\_poly/Al\_mesh & 6.3 MK \\
2010/01/30 & 15:13 & 2.2 & Ti\_poly/Al\_mesh & 1.6 MK \\
\enddata 
\label{tb:temp}
\end{deluxetable}

\begin{deluxetable}{cllcccccc} 
\tabletypesize{\footnotesize}
\tablecaption{Mass and size in X-ray passbands} 
\tablewidth{-100pt}
\tablehead{ \colhead{~Date} & \colhead{Time} & \colhead{Filter} & \colhead{S/N} &
\colhead{Depth/Width} & \colhead{Length} & \colhead{Volume} & \colhead{Electron density} & \colhead{Mass}  \\
   & (UT)  &        &           & (cm)                 & (cm)       & (cm$^3$)                & number (cm$^{-3}$)            & (g) } 
\startdata
2007/05/23         & 07:10    & Ti\_poly    & 3.4    & 2.7$\times$10$^9$    & 1.2$\times$10$^{10}$          & 6.9$\times$10$^{28}$ & 3.6$\times$10$^8$            & 4.8$\times$10$^{13}$            \\
2008/04/09         & 09:52    & Al\_poly    & 1.4$^\dag$    & 3.0$\times$10$^{9\ddag}$    & 3.4$\times$10$^{10\ddag}$    & 6.3$\times$10$^{29\flat}$ & 8.8$\times$10$^{7\dag}$ & 1.2$\times$10$^{14\flat}$  \\
2008/12/30         & 18:30    & Be\_thin    & 4.8    & 3.3$\times$10$^9$    & 1.7$\times$10$^{10}$          & 1.5$\times$10$^{29}$ & 1.3$\times$10$^8$            & 3.8$\times$10$^{13}$            \\
2009/01/11         & 23:21    & Be\_thin    & 3.4    & 5.0$\times$10$^9$    & 1.6$\times$10$^{10}$          & 3.1$\times$10$^{29}$  & 7.0$\times$10$^7$            & 4.5$\times$10$^{13}$            \\
2009/12/13         & 09:13    & Al\_med    & 4.1    & 3.6$\times$10$^9$    & 1.5$\times$10$^{10}$          & 1.5$\times$10$^{29}$  & 1.1$\times$10$^8$            & 3.3$\times$10$^{13}$            \\
2010/01/22         & 14:05    & Ti\_poly    & 2.4    & 2.6$\times$10$^9$     & 1.2$\times$10$^{10}$          & 6.4$\times$10$^{28}$  & 4.9$\times$10$^8$            & 6.3$\times$10$^{13}$ \\
2010/01/30         & 15:13    & Ti\_poly    & 1.5    & 5.3$\times$10$^9$     & 3.1$\times$10$^{10}$          & 6.8$\times$10$^{29}$  & 3.9$\times$10$^8$            & 5.4$\times$10$^{14}$           \\
\enddata 
\label{tb:mass}
\tablecomments{\\
$^\dag$ The S/N and density are the average values of the region A and B. \\
$^\ddag$ The depth/width and length are the values from B region. \\
$^\flat$ The volume and mass are the summation of the region A and B.}
\end{deluxetable}

\begin{deluxetable}{ccccccccccccc} 
\tabletypesize{\scriptsize}
\tablecaption{Energies and time scales of erupting X-ray plasma}
\tablewidth{-100pt}
\tablehead{ \colhead{~Date} & \colhead{${\tau_{rad}}$} & \colhead{${\tau_{cond}}$} & \colhead{$\bigtriangleup$t} &&
\colhead{L$_r$} &  \colhead{F$_c$} & \colhead{T.E.} & \colhead{Time$^a$} & \colhead{V$^b$} & \colhead{K.E.} & r$^c$ & \colhead{P.E.$^d$}  \\ \cline{2-4} \cline{6-8}
  & \multicolumn{3}{c}{(10$^3$s)}  & & \multicolumn{3}{c} {(10$^{28}$erg)} & (UT) & (km/s) & (10$^{28}$erg) & (r$_{\sun}$) & (10$^{28}$erg) }

\startdata
2007/05/23 & 320    & 0.044  &  3.8     &&  0.14 & 800 &  18 & 07:00$-$07:12 & 50 & 0.06          & ~1.16                  & ~~1.2   \\
2008/04/09 & 920    & 0.15    &  1.1     &&  0.026 & 46$^\dag$ &  36$^\ddag$  & 09:40$-$09:50 & 72 & 0.32$^\ddag$  & $>$1.23$^\flat$  & $>$4.2    \\            
2008/12/30 & 89      & 1.0       &  15      &&  0.39 & 27 & 3.5 & 18:31$-$19:53 & ~5 & 0.0005         & ~1.31                  & ~~1.7   \\
2009/01/11 & 170    & 1.9       &  1.8     &&  0.030 & 3.9 & 4.1 & 22:50$-$23:21 & 35 & 0.028        & ~1.26                  & ~~1.8     \\
2009/12/13 & 1100  & 0.021  &  0.30   &&  0.002 & 91 & 13 & 09:08$-$09:19 & 70 & 0.081         & ~1.28                  & ~~1.4    \\
2010/01/22 & 110     & 0.75        &  0.84   & & 0.07 & 16.3  & 15 & 13:53$-$14:08 & 45 & 0.06            & $>$1.20$^\flat$   & $>$1.9      \\
2010/01/30 & 14      & 31        &  1.7     &&  2.5 & 0.87 & 32 & 14:45$-$15:10 & 20 & 0.11          & $>$1.17$^\flat$   & $>$15       \\
\enddata 
\label{tb:energy}
\tablecomments{\\
$^a$ Time used for the velocity estimation (UT) \\
$^b$ Velocity of eruptive plasma in X-ray (km/sec) \\
$^c$ Final height of eruptive plasma (R$_\sun$) \\
$^d$ Potential energy\\
$\dag$ Thermal conduction is estimated using the loop structure of the region B. \\
$\ddag$ T.E. and kinetic energy (K.E.) are the summation of both A and B regions. \\
$\flat$ The highest height that can be measured in the limited FOV from the observations. }
\end{deluxetable}

\begin{figure}
\epsscale{1.5}\plotone{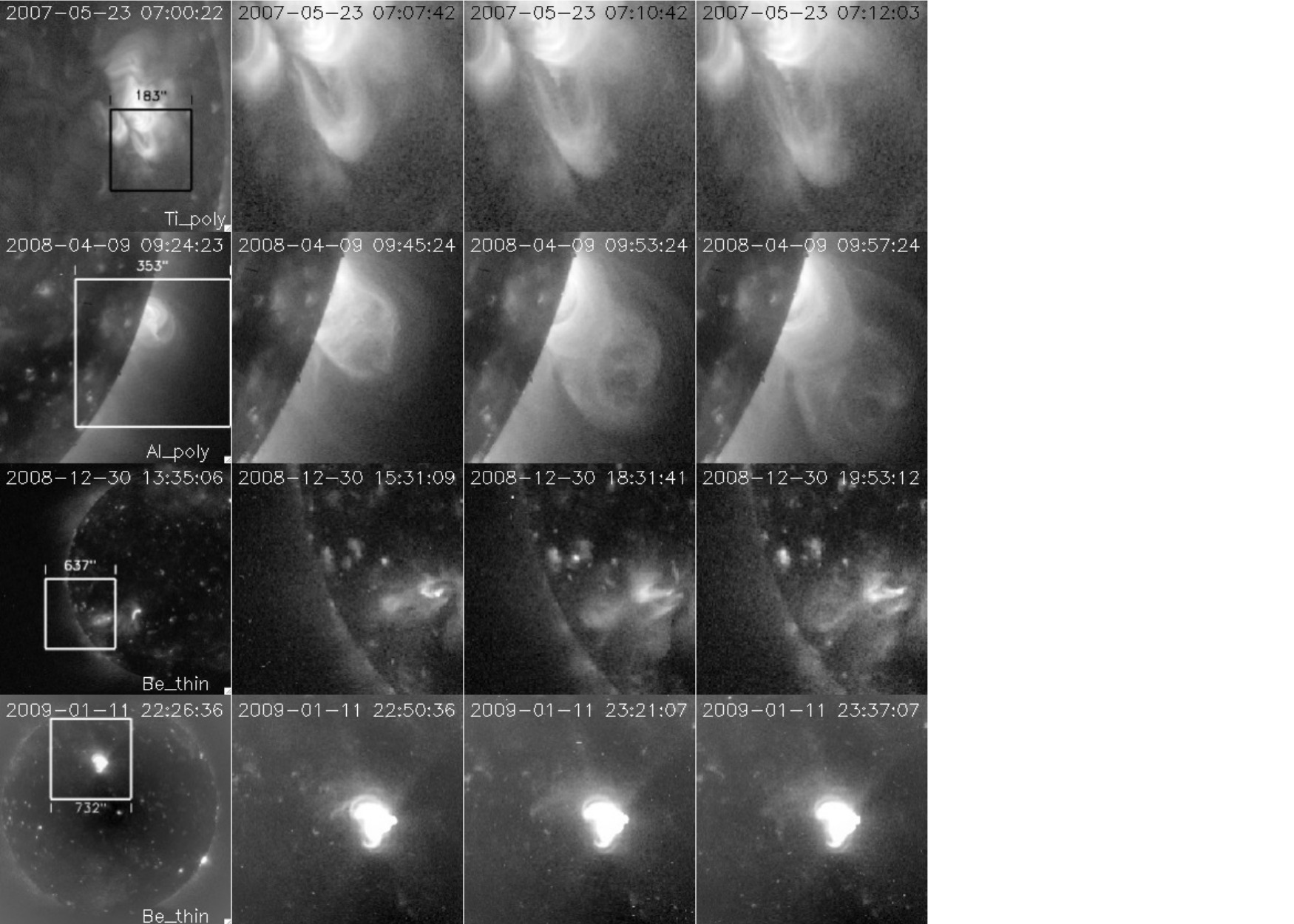}
\caption{Time series of images of seven eruptive plasmas. 
The first column shows the observations in the observed FOV before the eruptions. The other columns show the eruptive plasmas in the small FOV of the box at the first columns at different times. } 
\label{fig:xrt}
\end{figure}
\addtocounter{figure}{-1}
\begin{figure}
\epsscale{1.1}\plotone{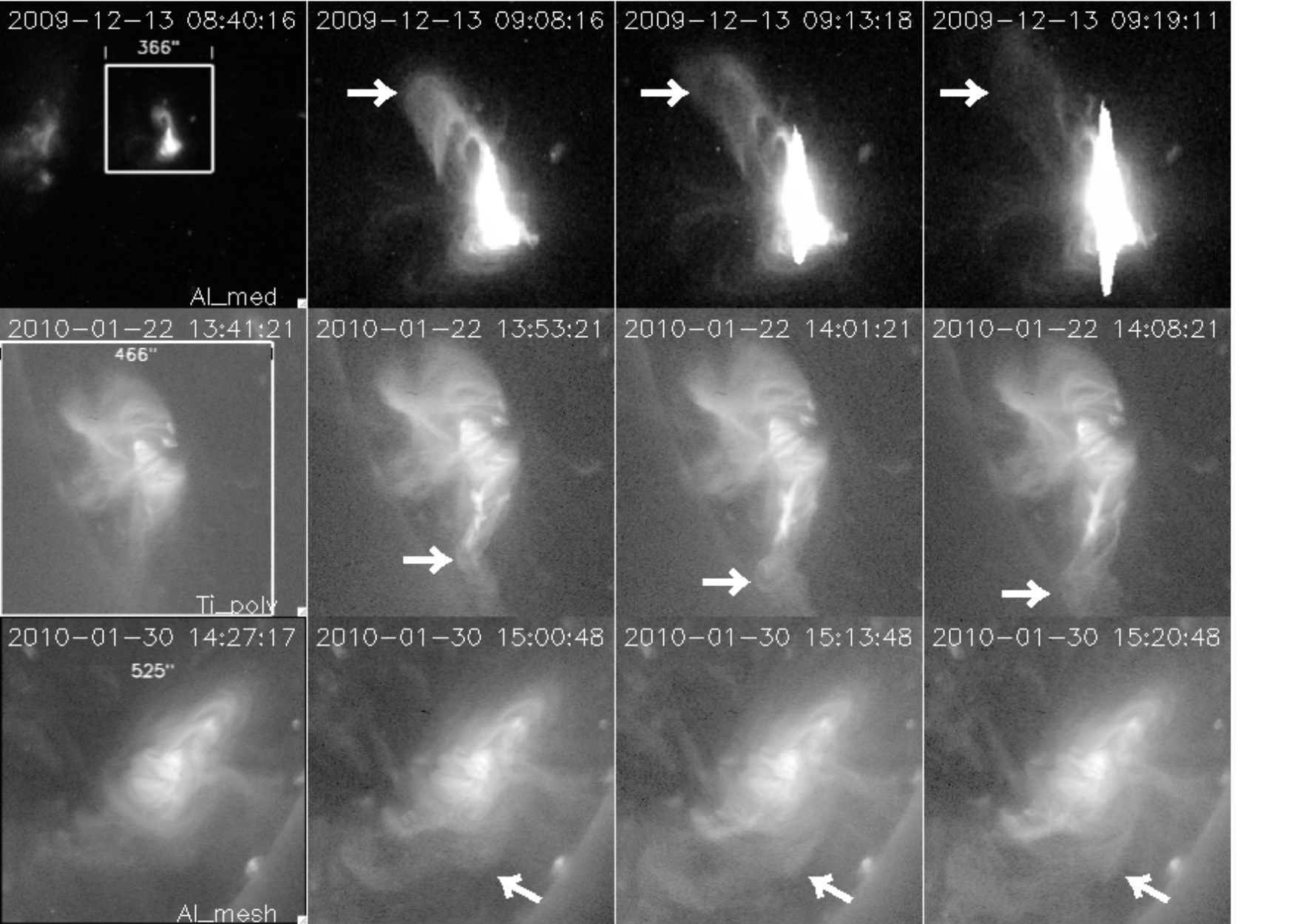}
\caption{(continued)} 
\label{fig:xrt}
\end{figure}

\begin{figure}
\epsscale{1.6}\plotone{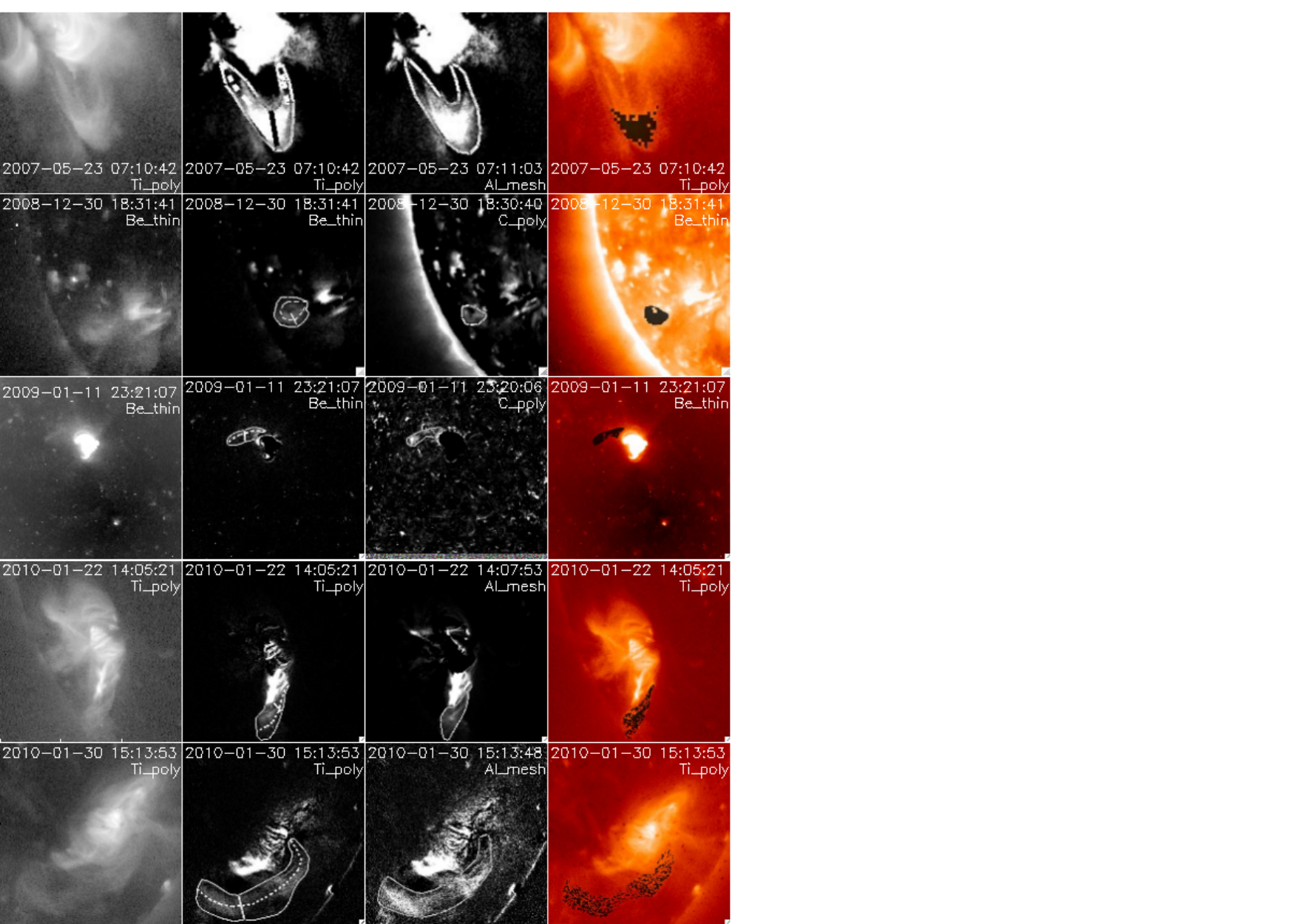}
\caption{Regions for the temperature and mass estimations. 
The first column shows the observations used in the temperature and mass estimations. 
The second and third columns show the regions of eruptive plasma on two passbands outlined by contoured lines. 
Solid and dashed lines inside the contoured regions in the second column represent 
the widths and lengths of the cylinder structure. 
The last column shows the selected pixels for the temperature estimation 
within the thresholds by a black color on the observations.} 
\label{fig:temp}
\end{figure}

\begin{figure}
\begin{tabular}{cc}
\includegraphics[width=70mm]{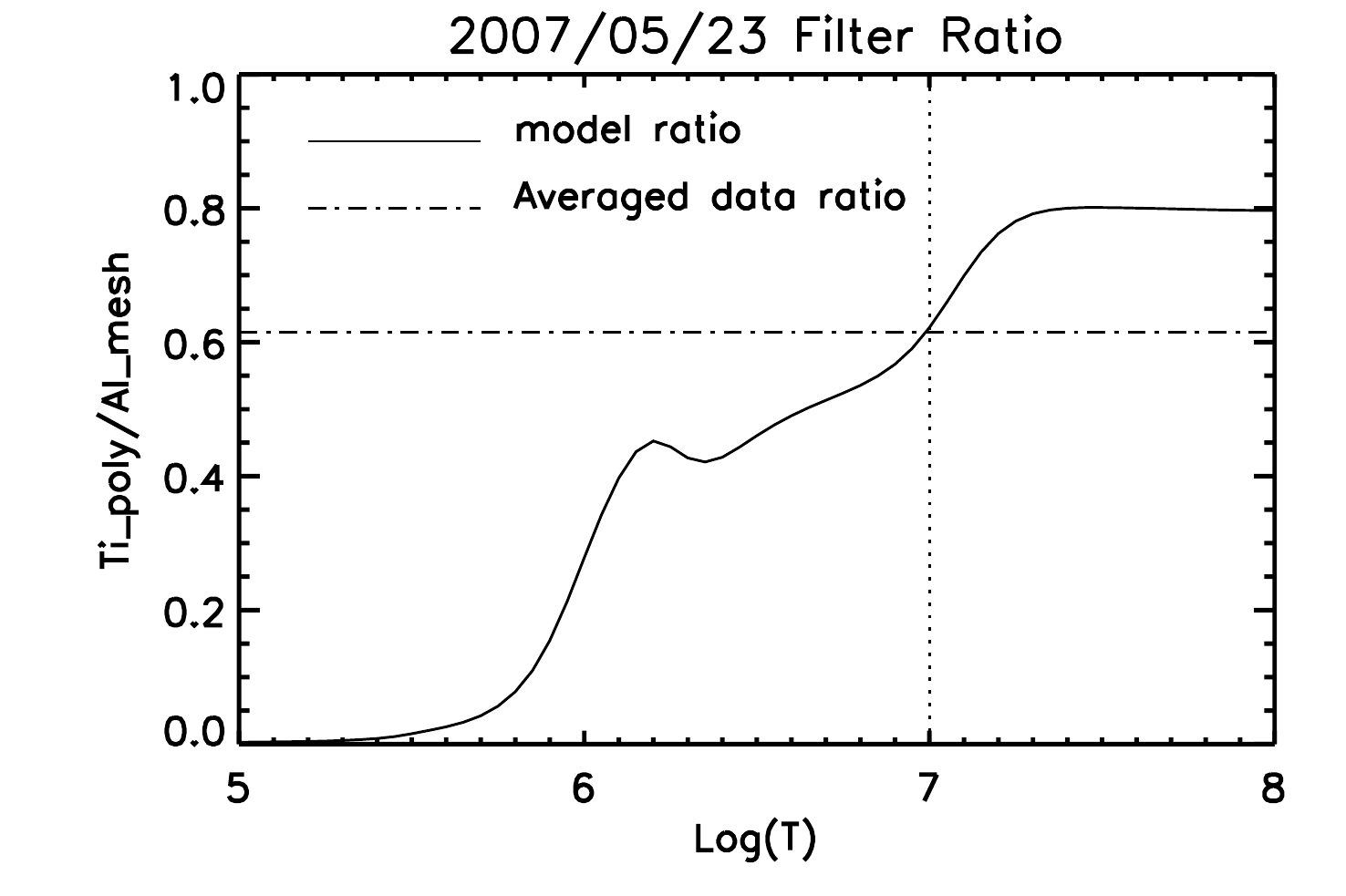} &
\includegraphics[width=70mm]{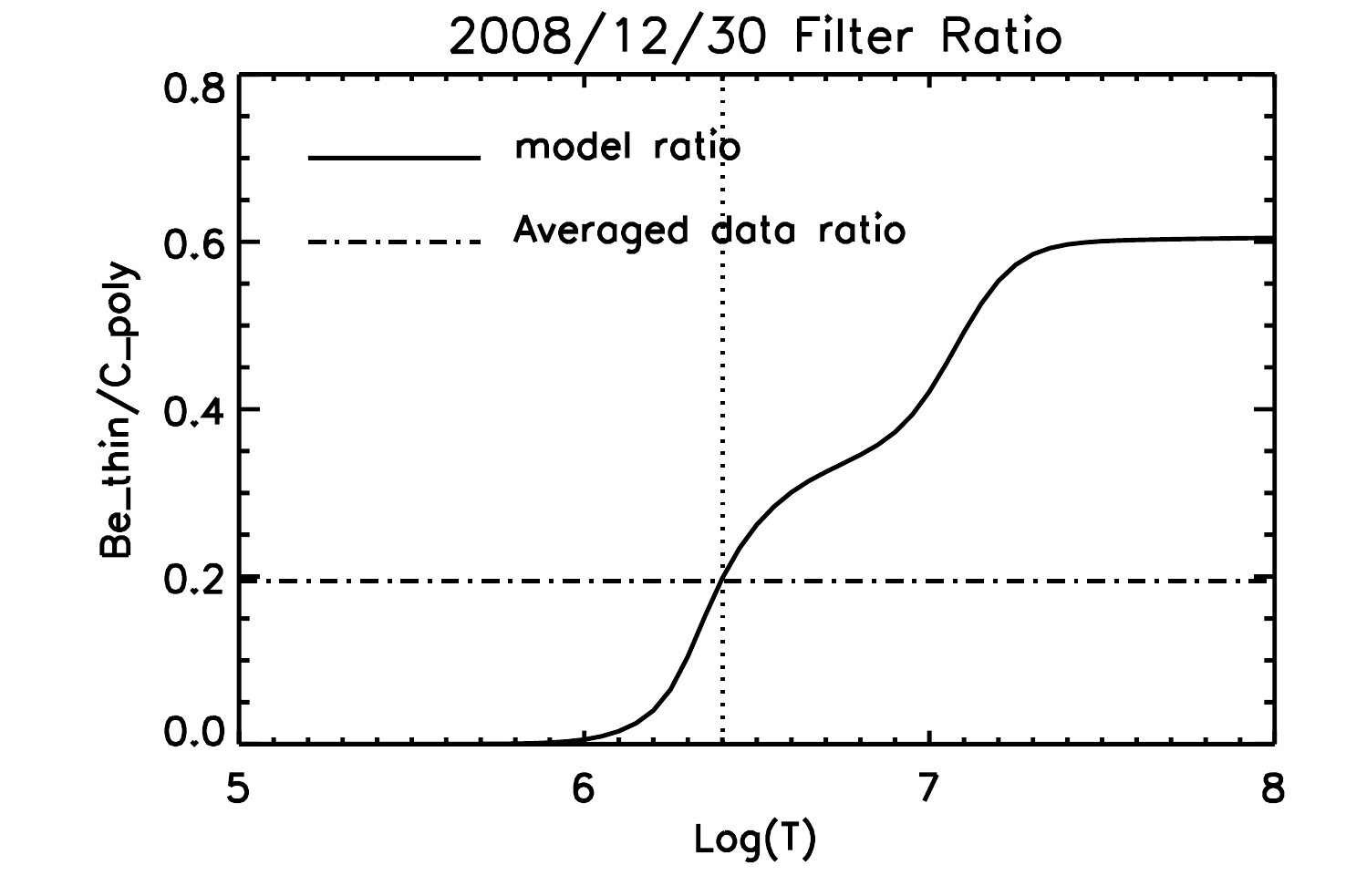} \\
\includegraphics[width=70mm]{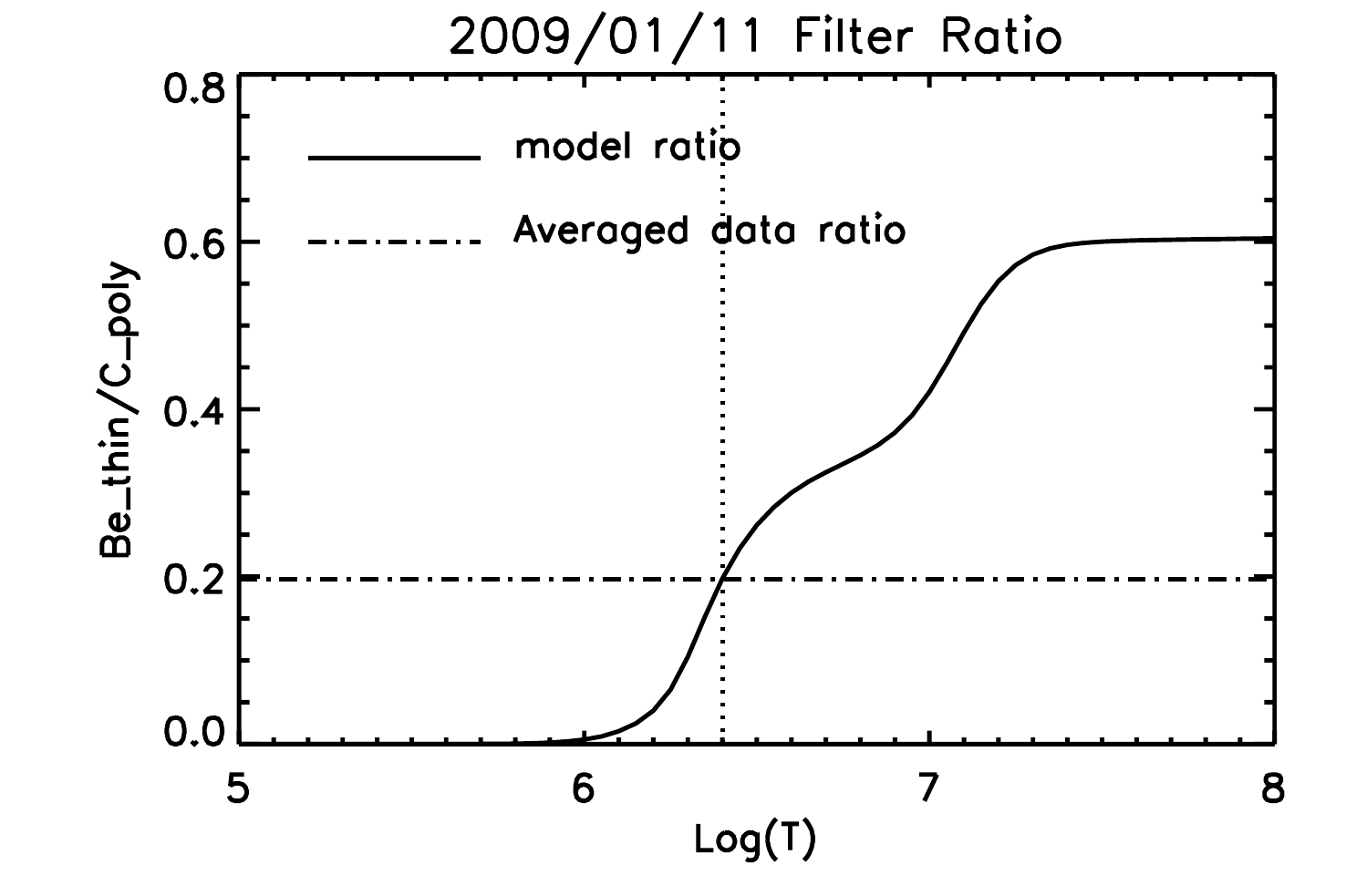} &
\includegraphics[width=70mm]{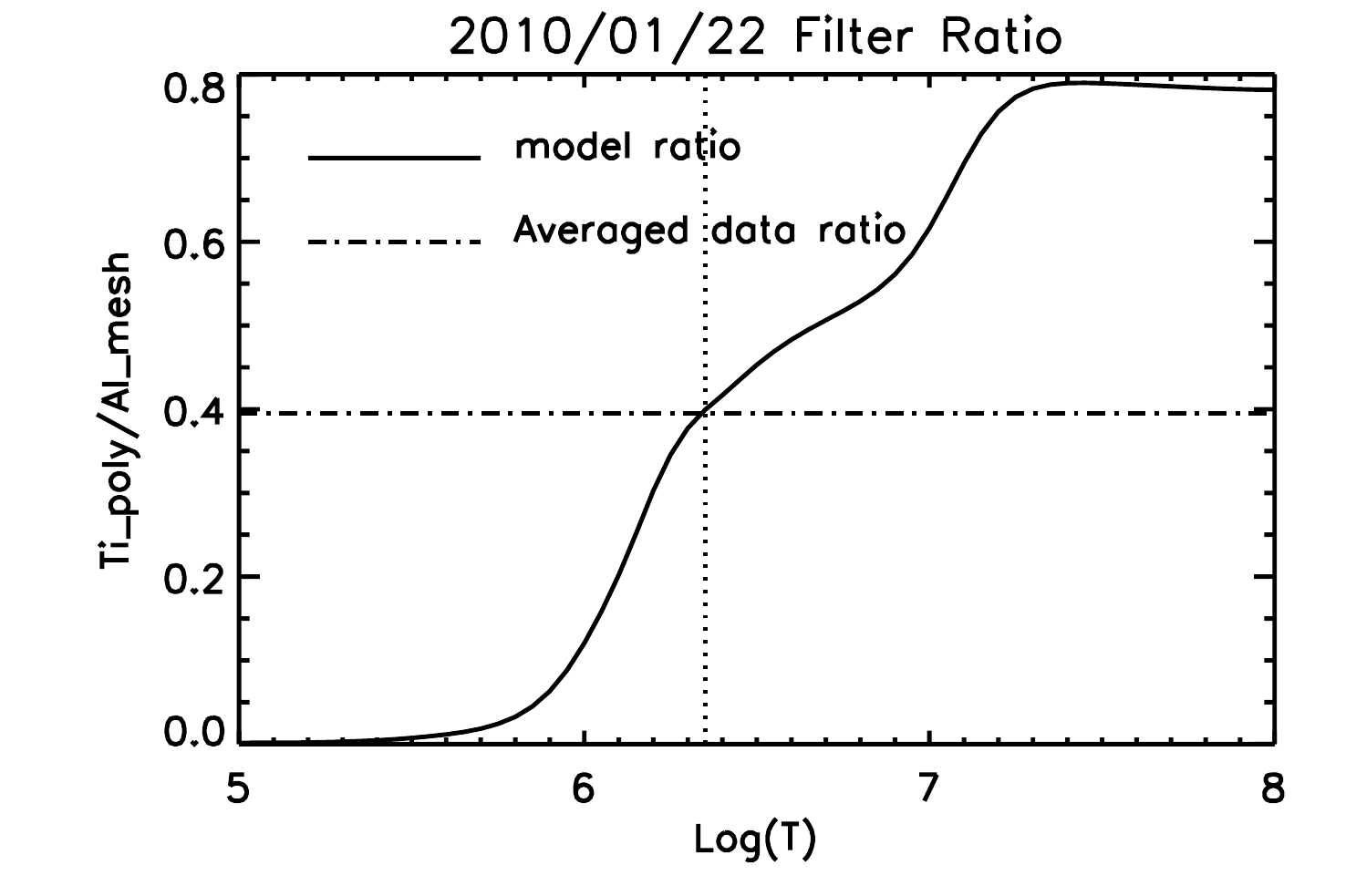} \\
\includegraphics[width=70mm]{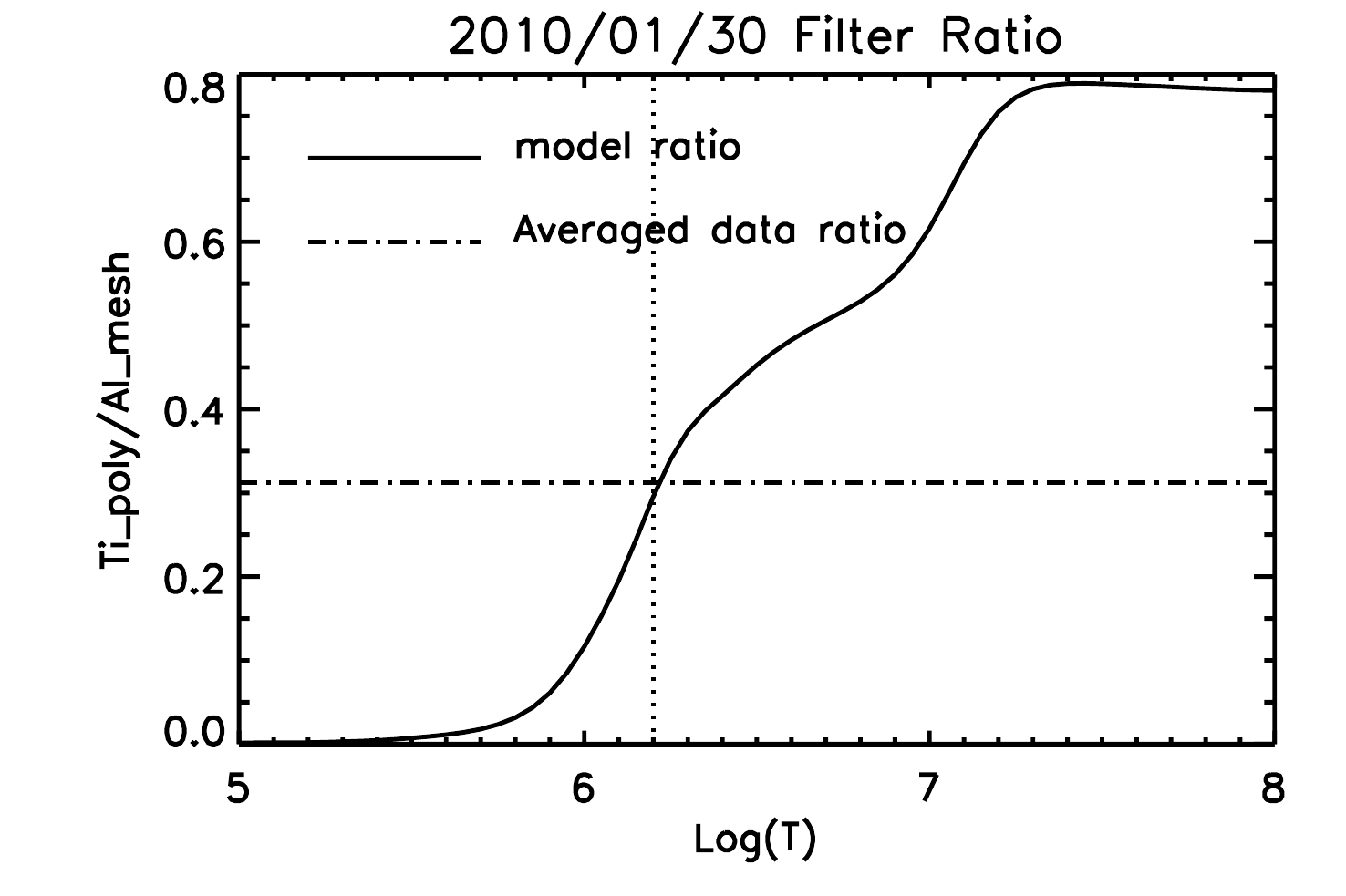} 
\end{tabular}
\caption{Averaged data ratios and estimated temperatures (dotted lines) for five events with the filter ratios by model.} 
\label{fig:ratio}
\end{figure}

\begin{figure}
\epsscale{1}\plotone{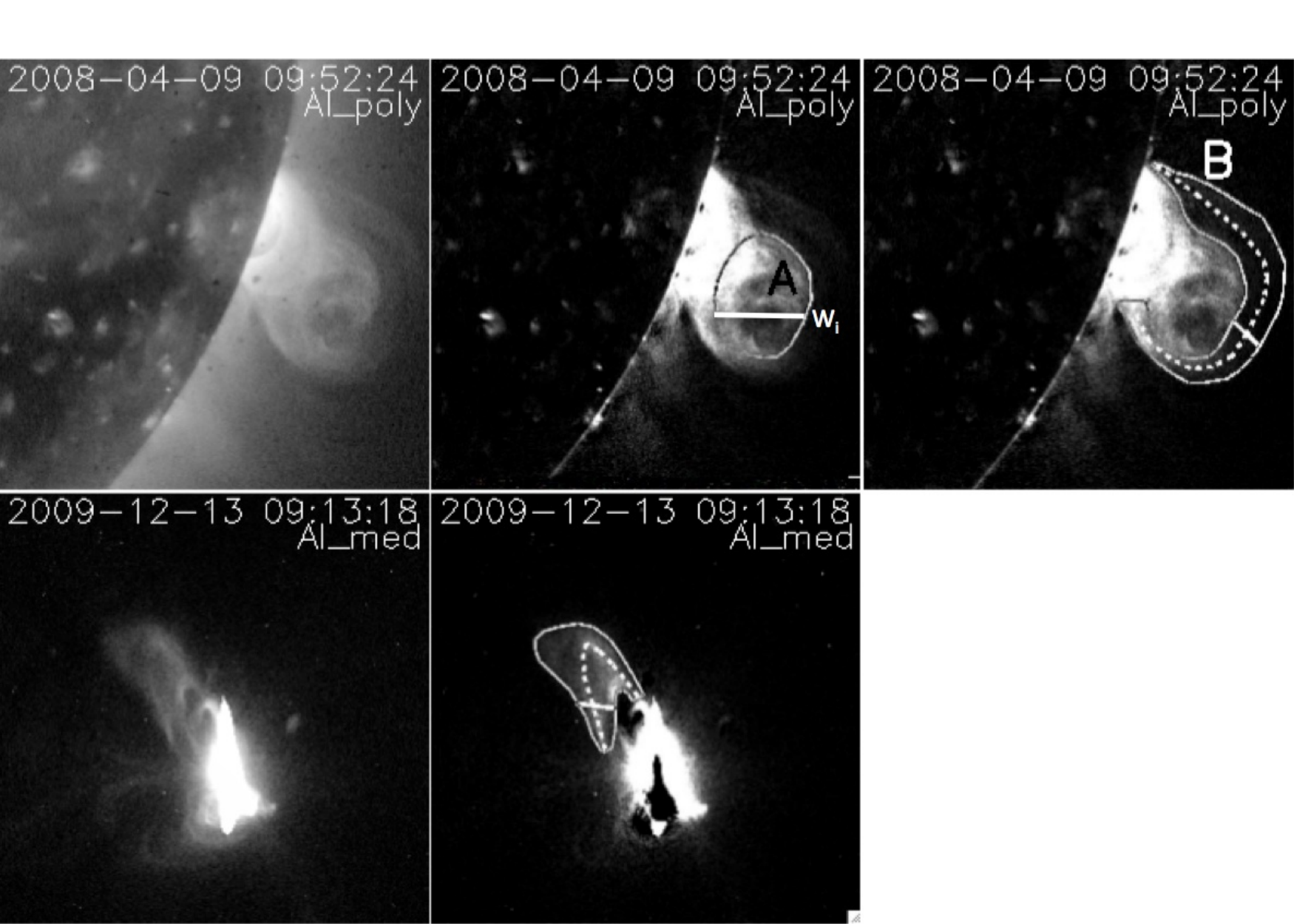}
\caption{Regions for the mass estimation for two events. 
The first column shows the observations used for the mass estimation. 
The other columns show the selected regions by the enclosed solid lines on the observations. 
The dashed and solid lines represent the length and width of the loop by the assumption of the structures, respectively. 
For the event on 2008. April 09, the two regions (A and B) are shown at the second and third columns in the top panel.
In the second column in the top panel, the solid line represents the width (w$_{\rm i}$) at a row (i) in the image (see Section~\ref{sec:mass}).
} 
\label{fig:mass}
\end{figure}

\begin{figure}
\begin{tabular}{cc}
\includegraphics[width=70mm]{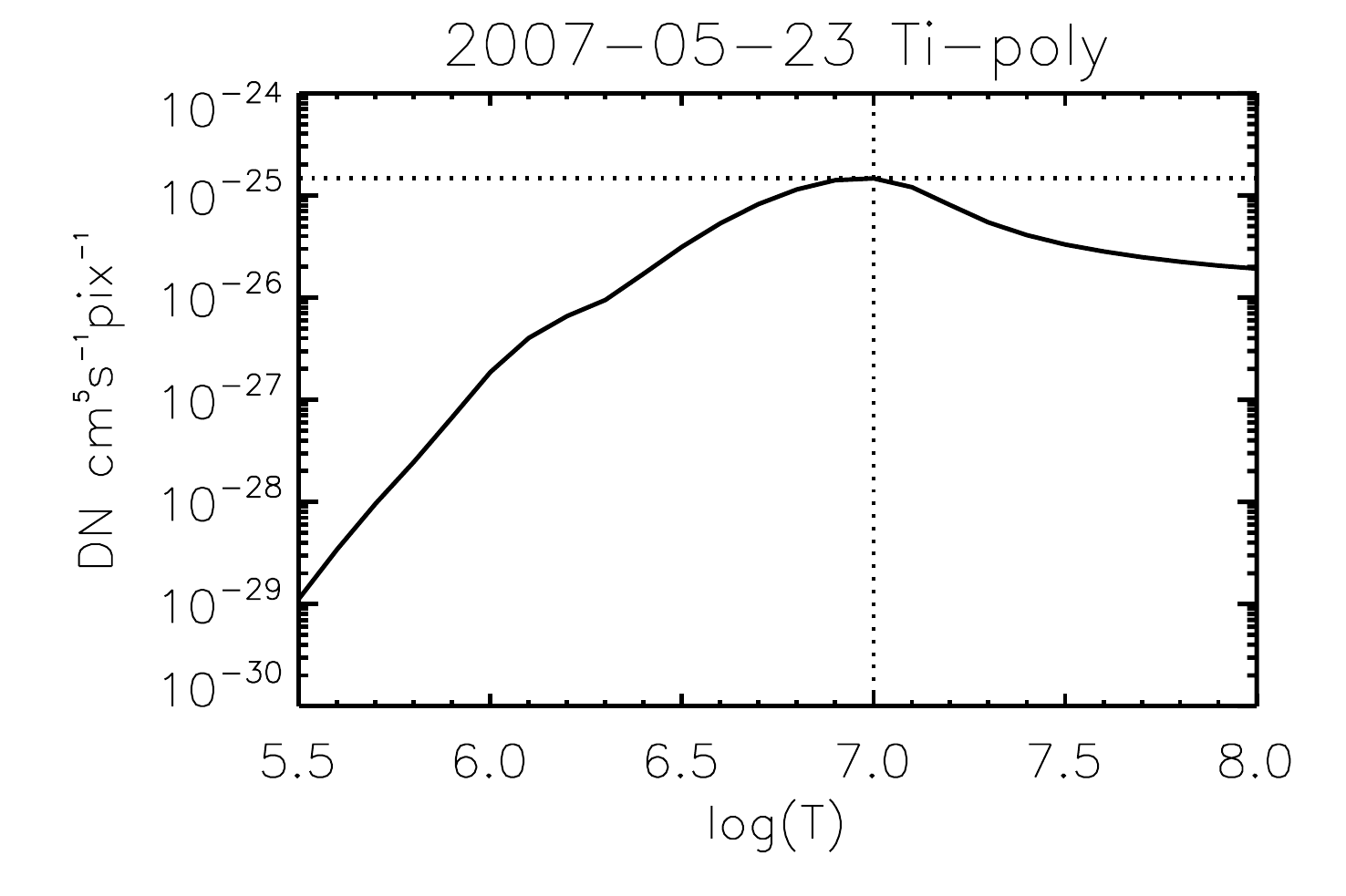} &
\includegraphics[width=70mm]{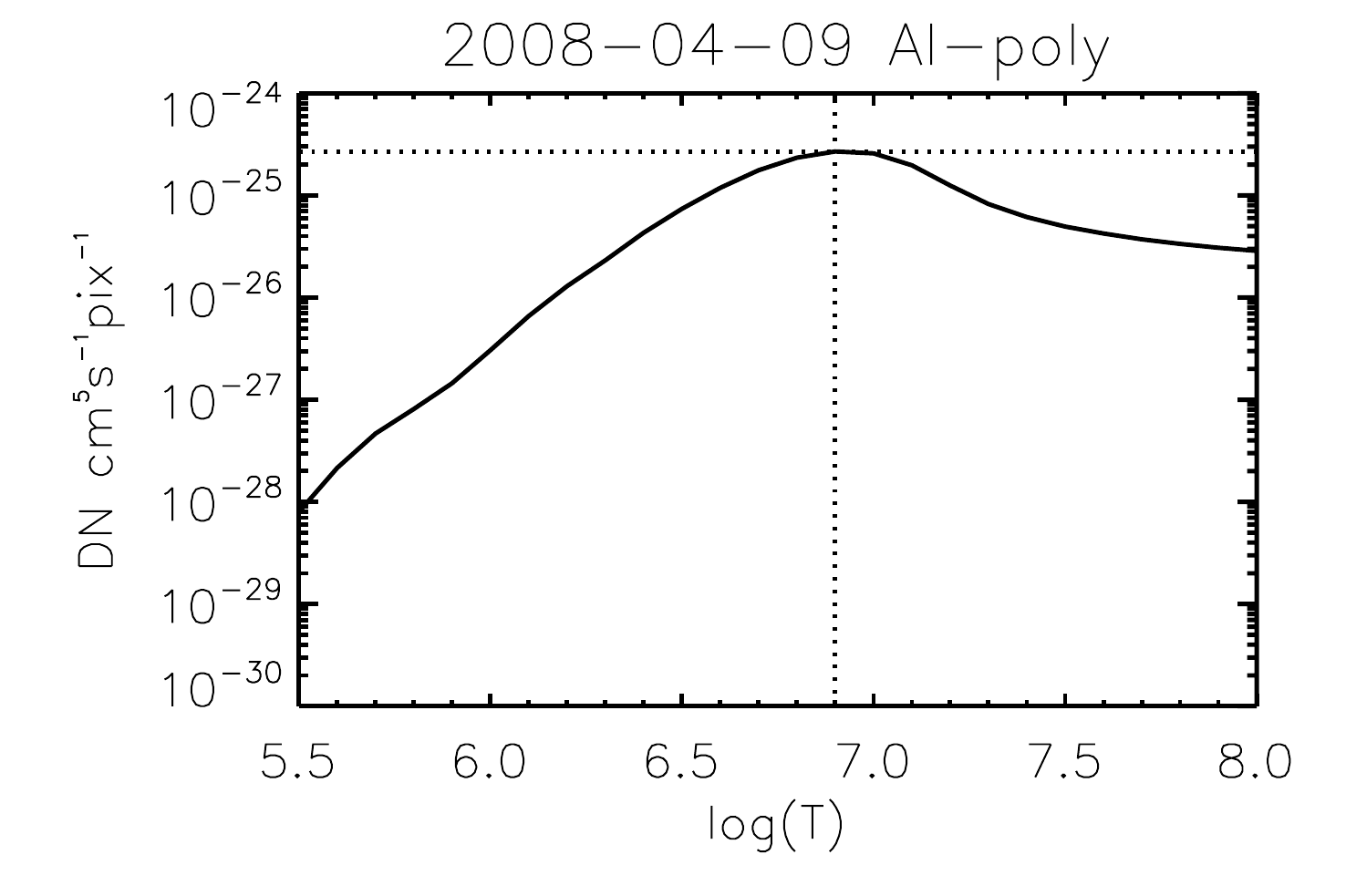} \\
\includegraphics[width=70mm]{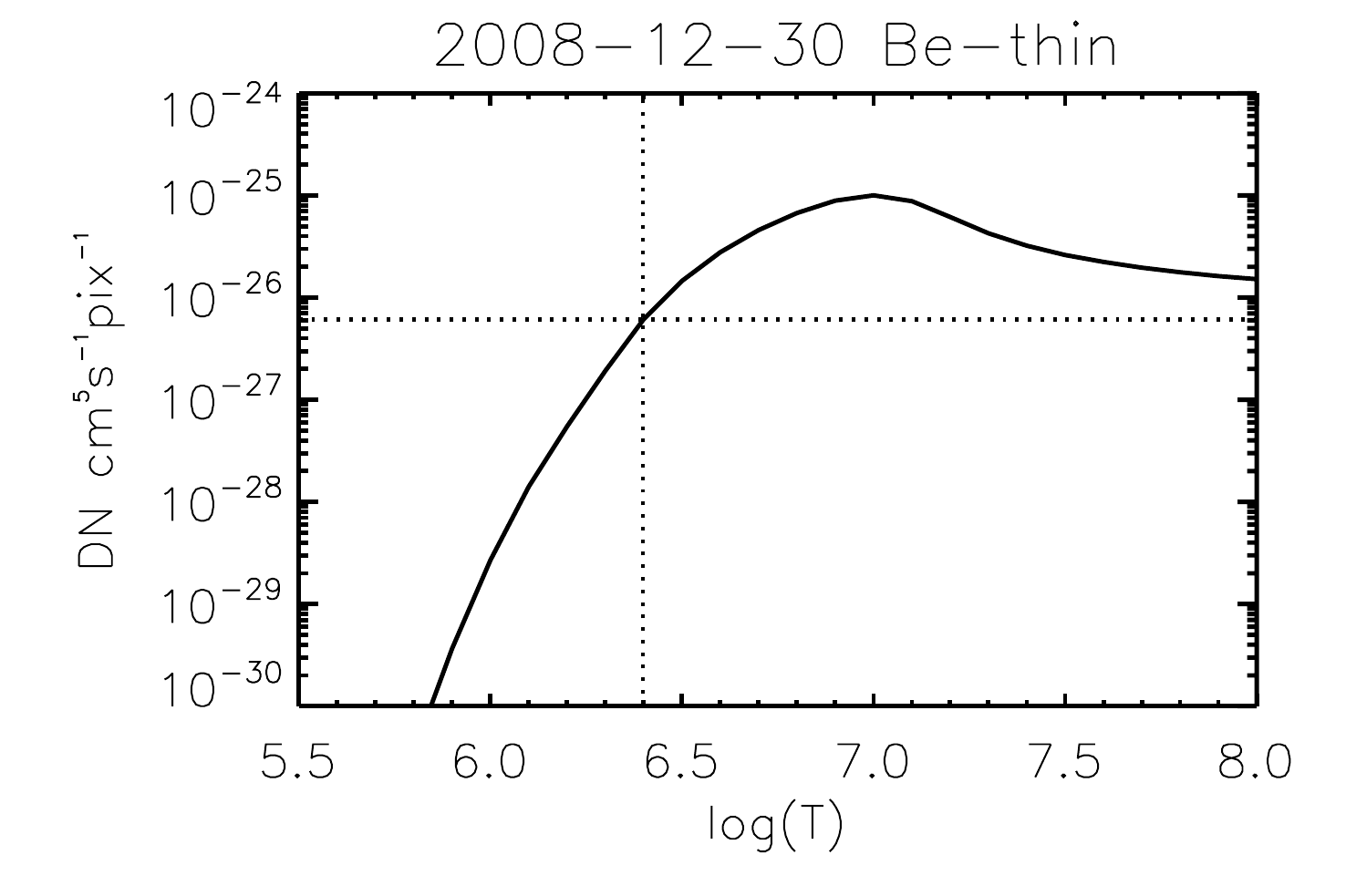} &
\includegraphics[width=70mm]{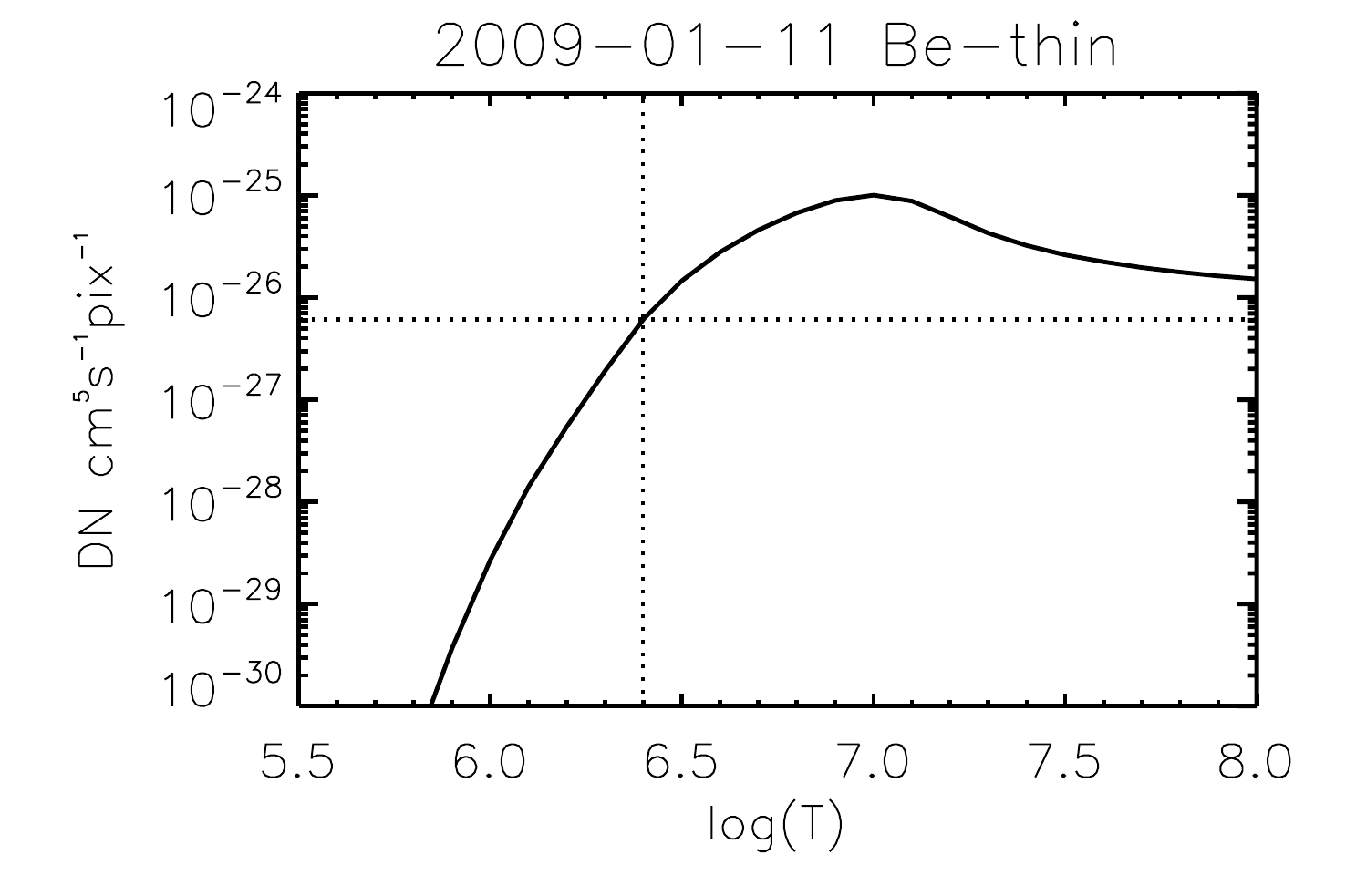} \\
\includegraphics[width=70mm]{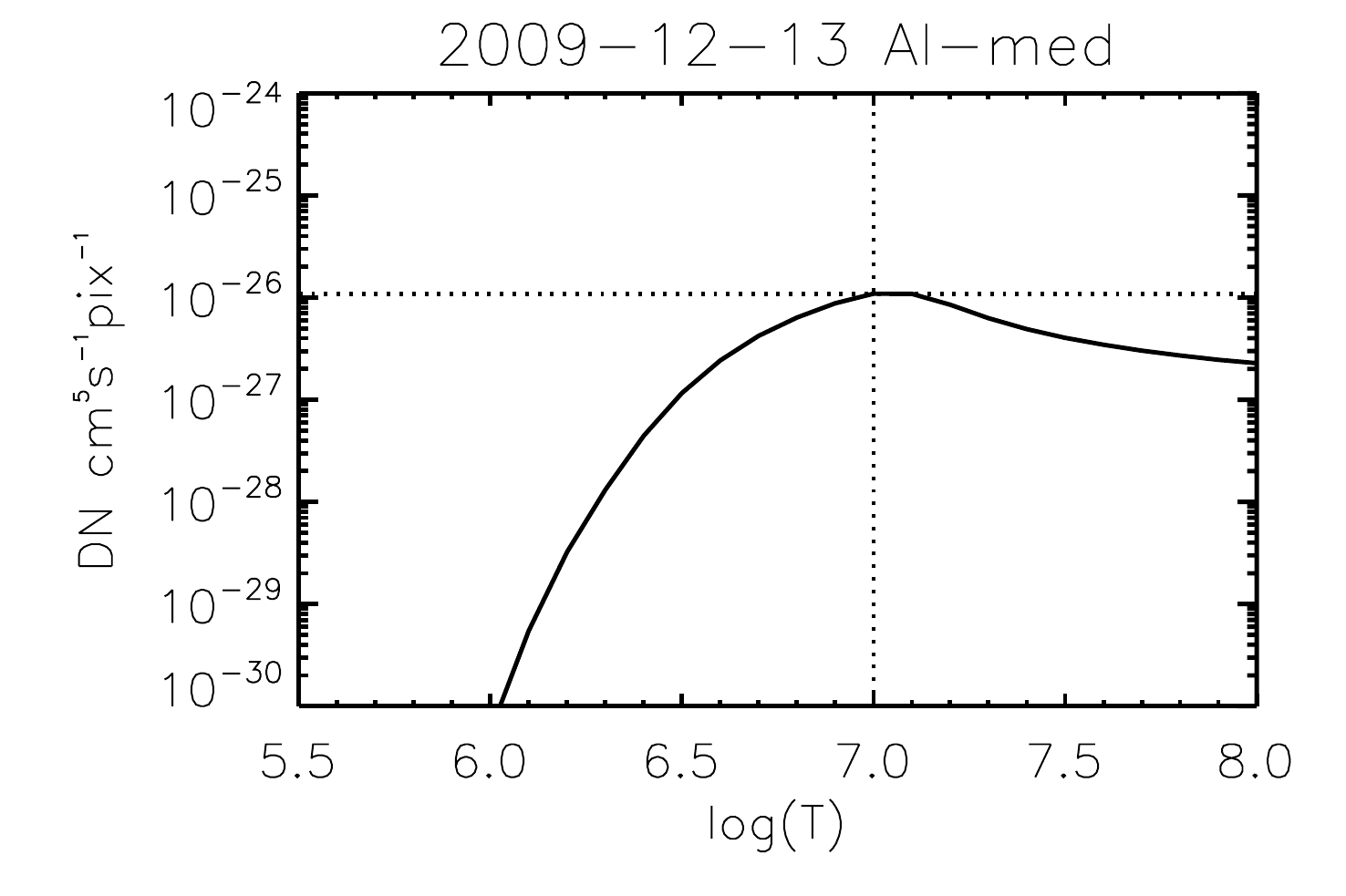} &
\includegraphics[width=70mm]{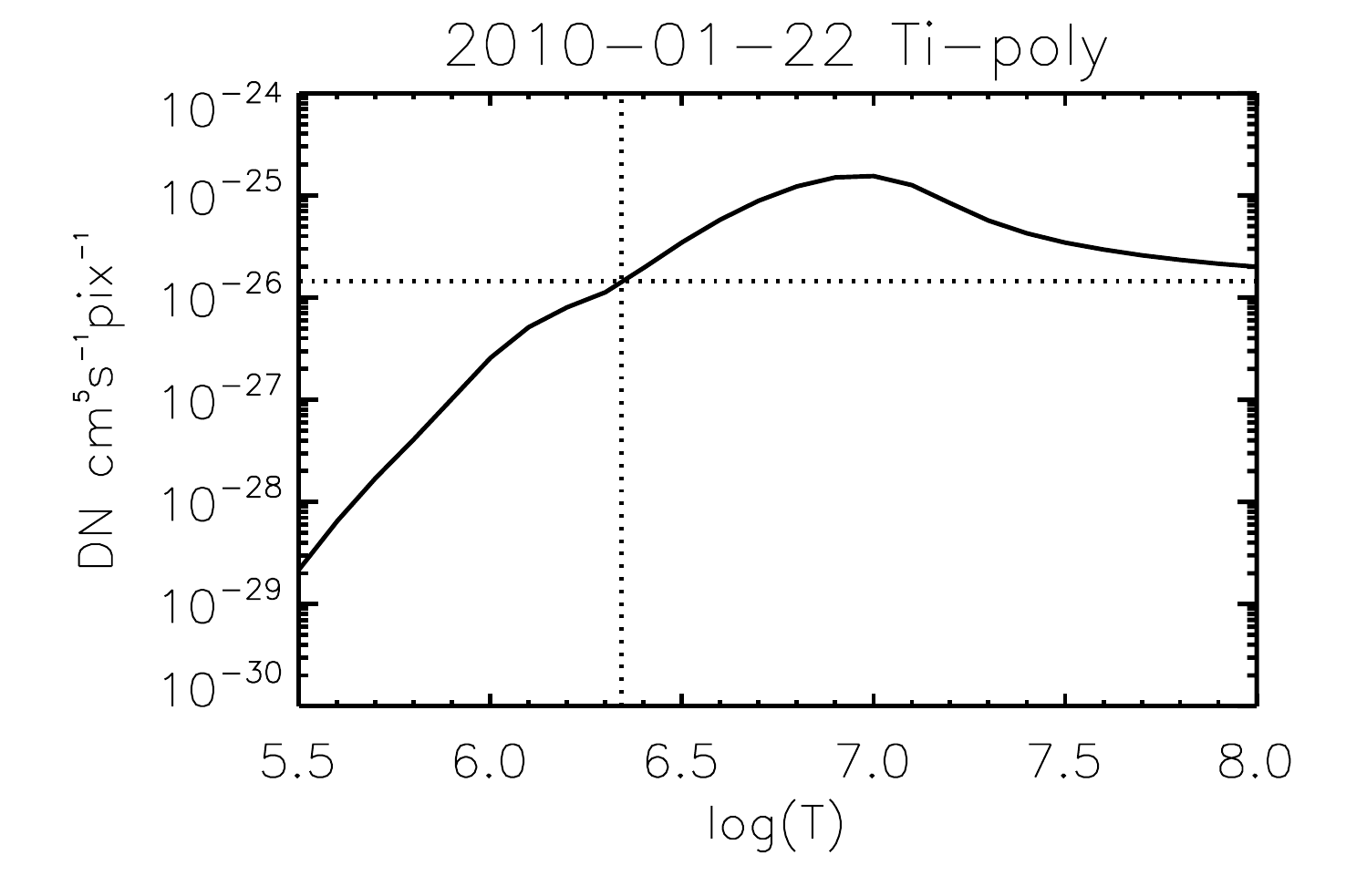} \\
\includegraphics[width=70mm]{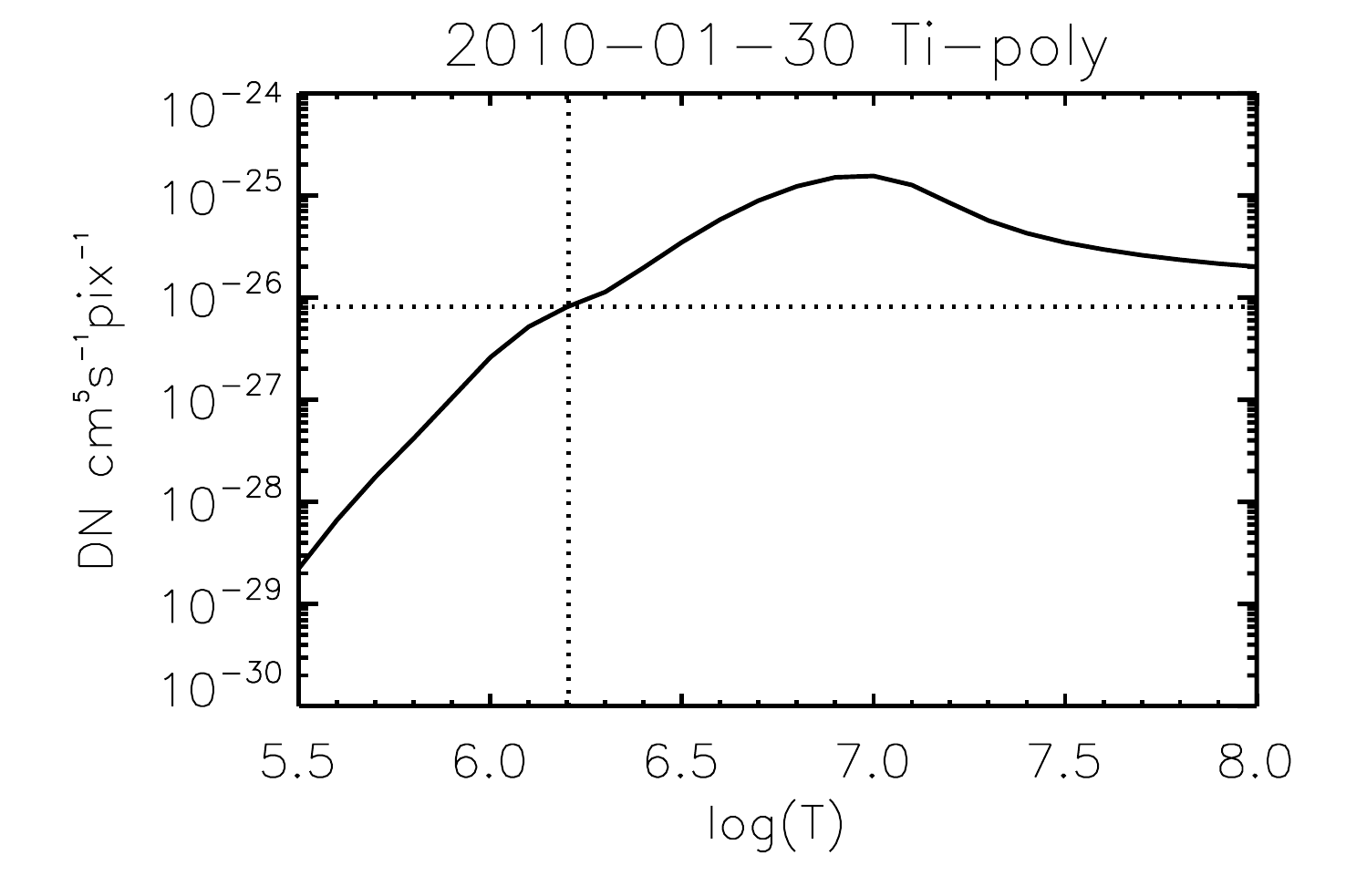} &
\end{tabular}
\caption{Temperature responses at the estimated temperatures for the estimation of emission measure.} 
\label{fig:resp}
\end{figure}

\begin{figure}
\begin{tabular}{cc}
\includegraphics[width=70mm]{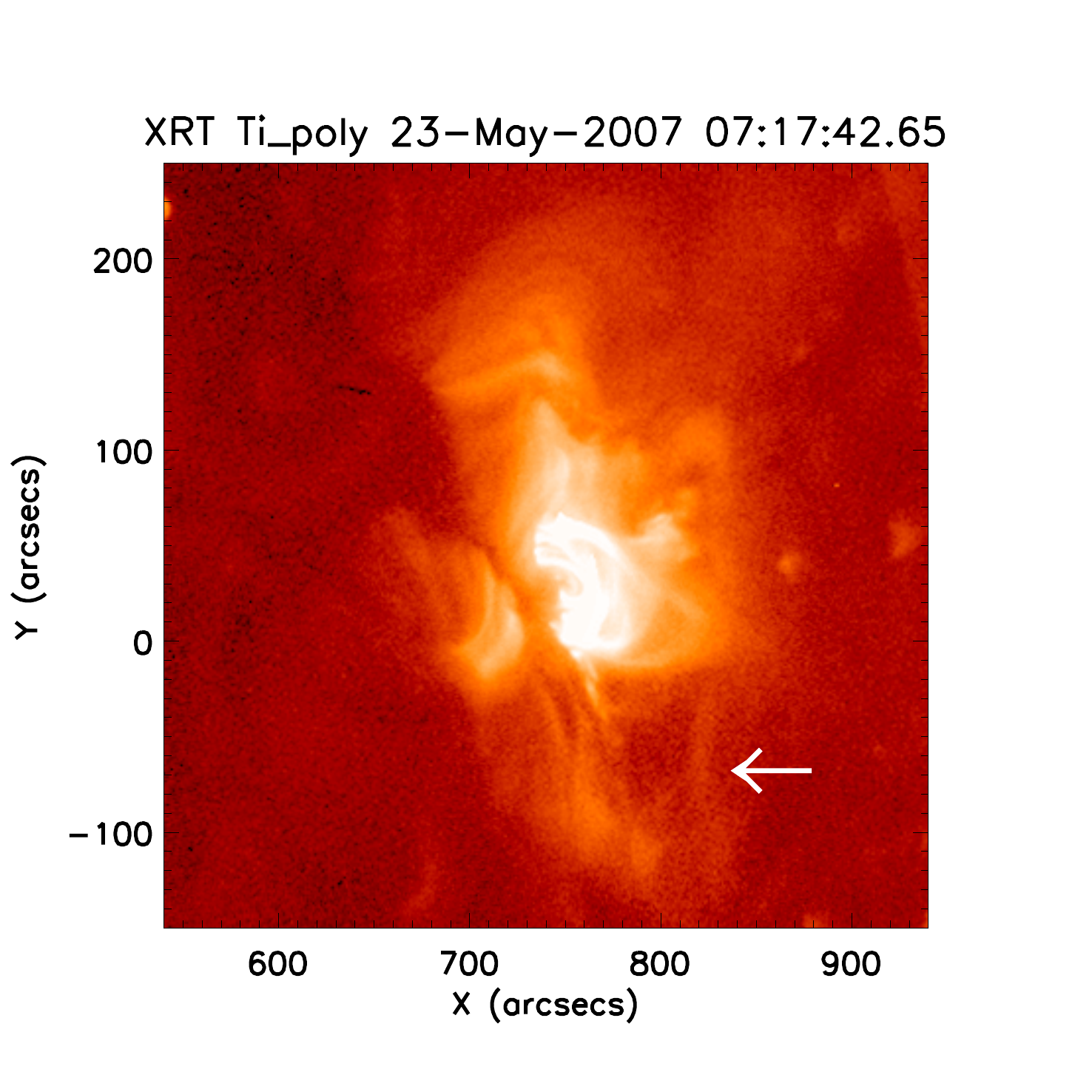} &
\includegraphics[width=70mm]{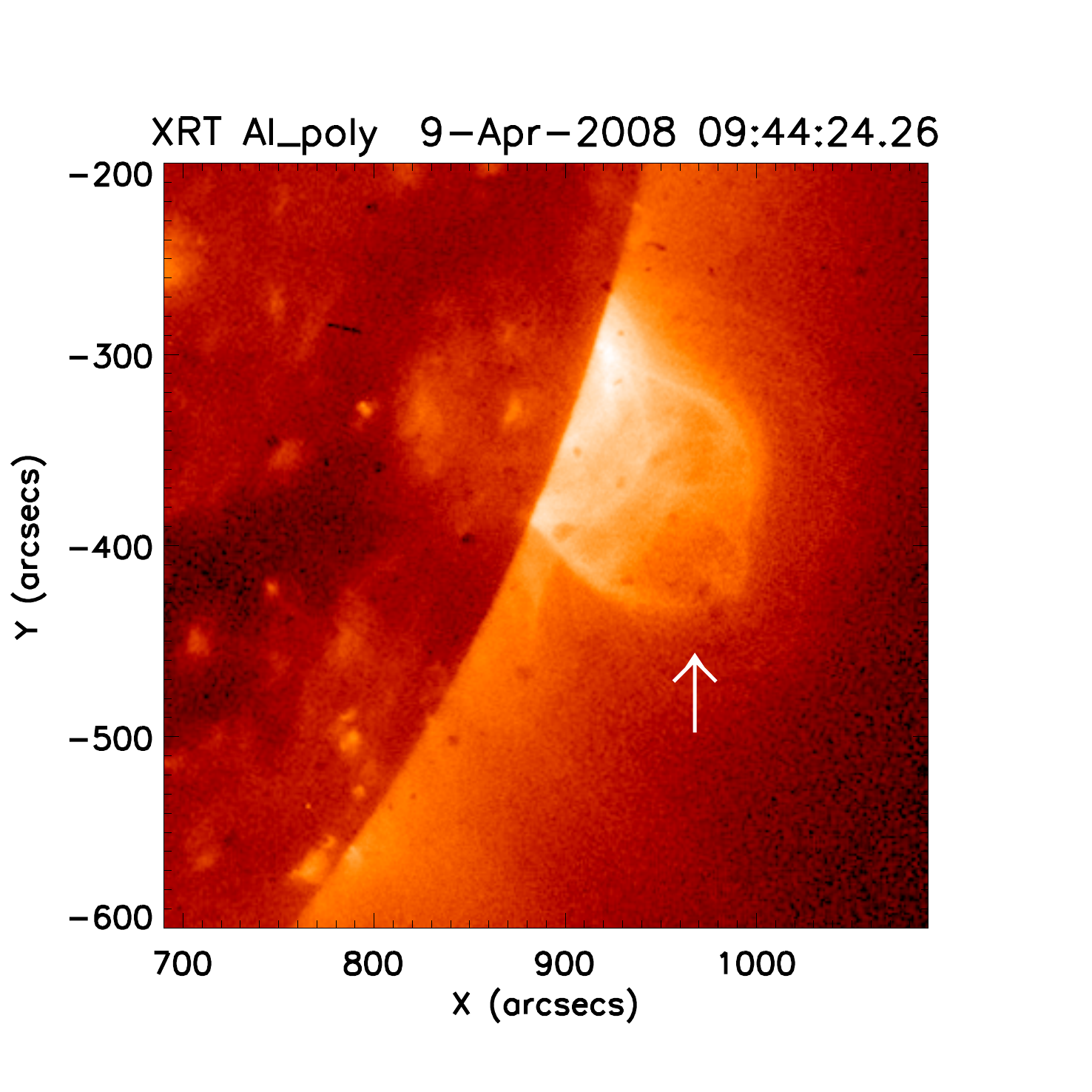} \\
\includegraphics[width=70mm]{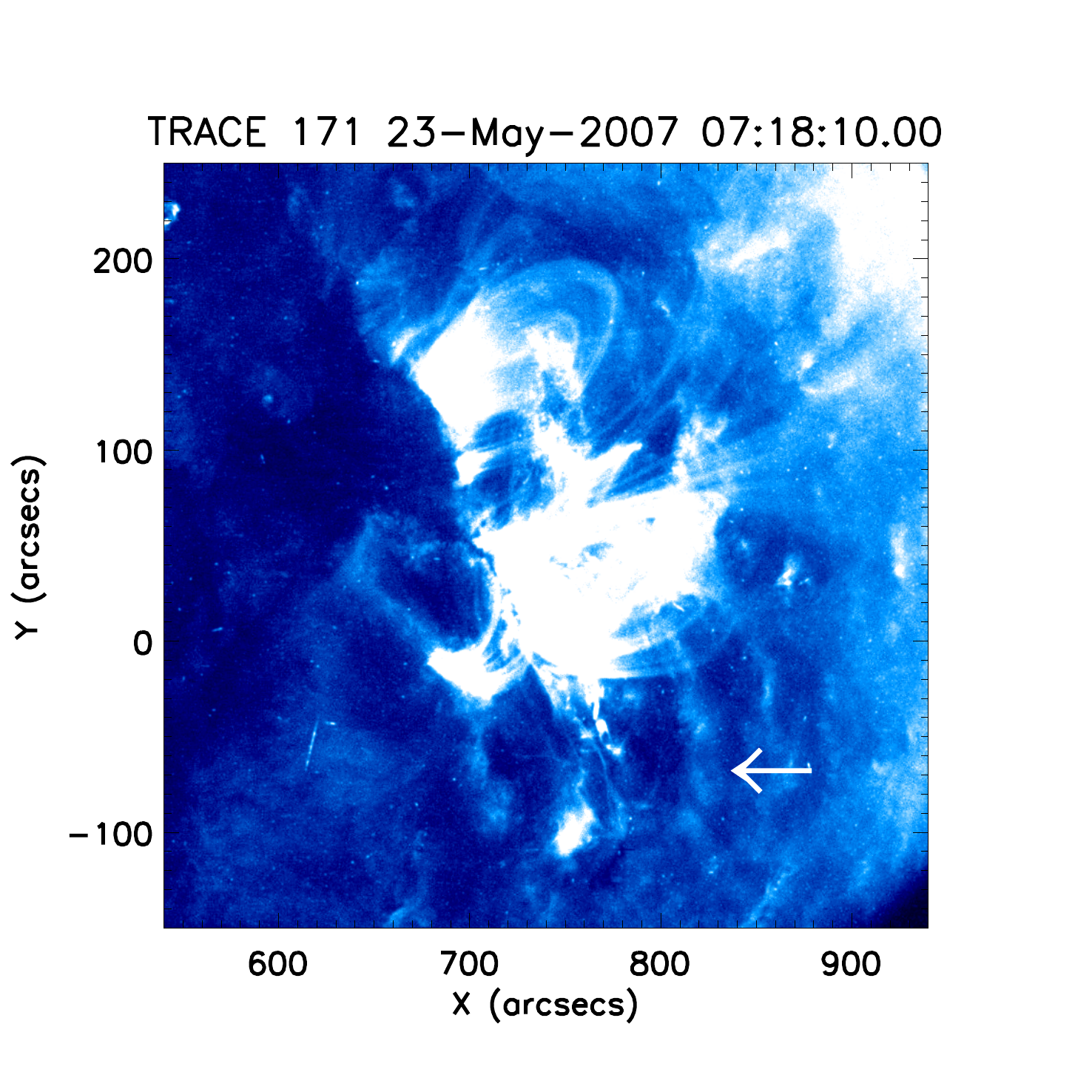} &
\includegraphics[width=70mm]{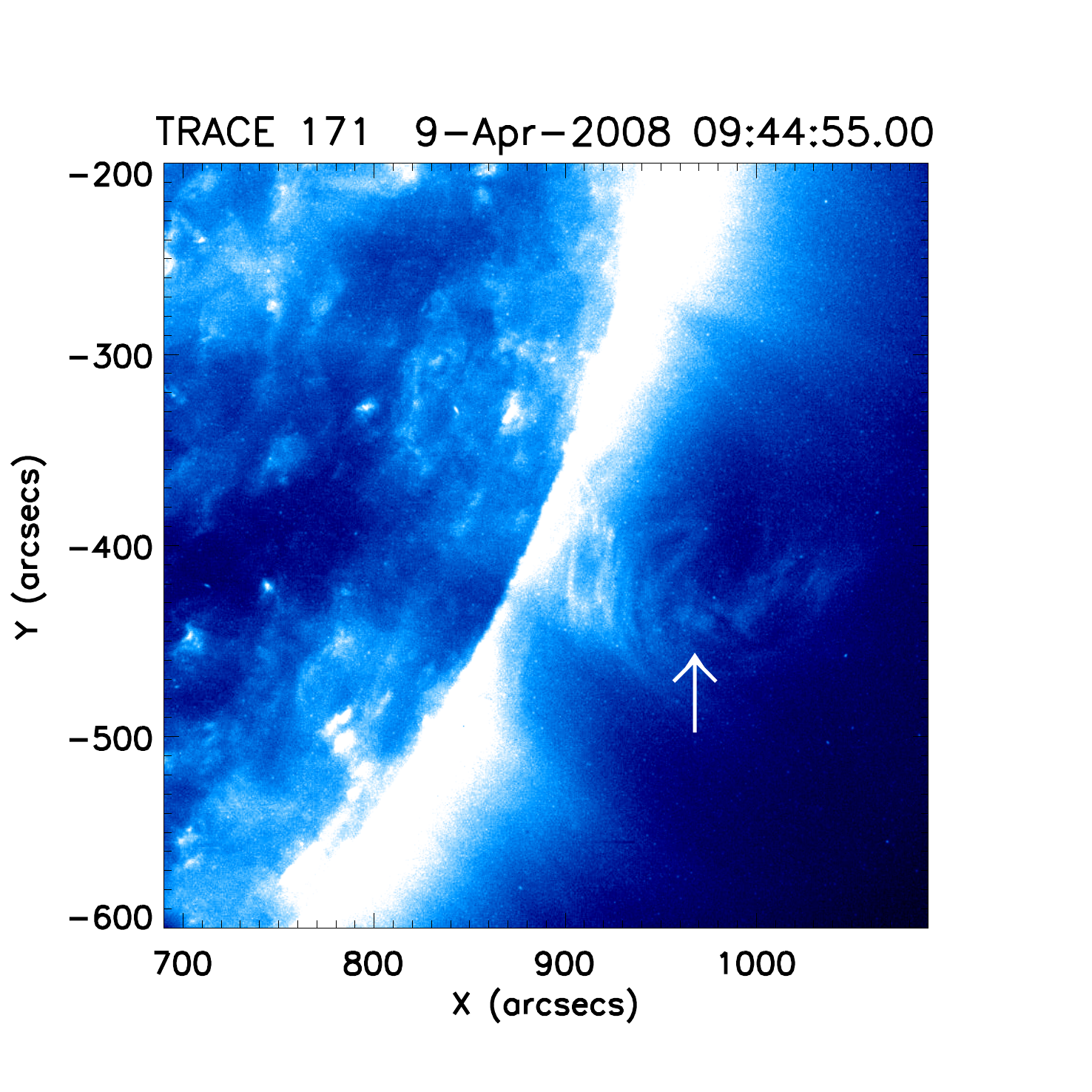} \\
\includegraphics[width=70mm]{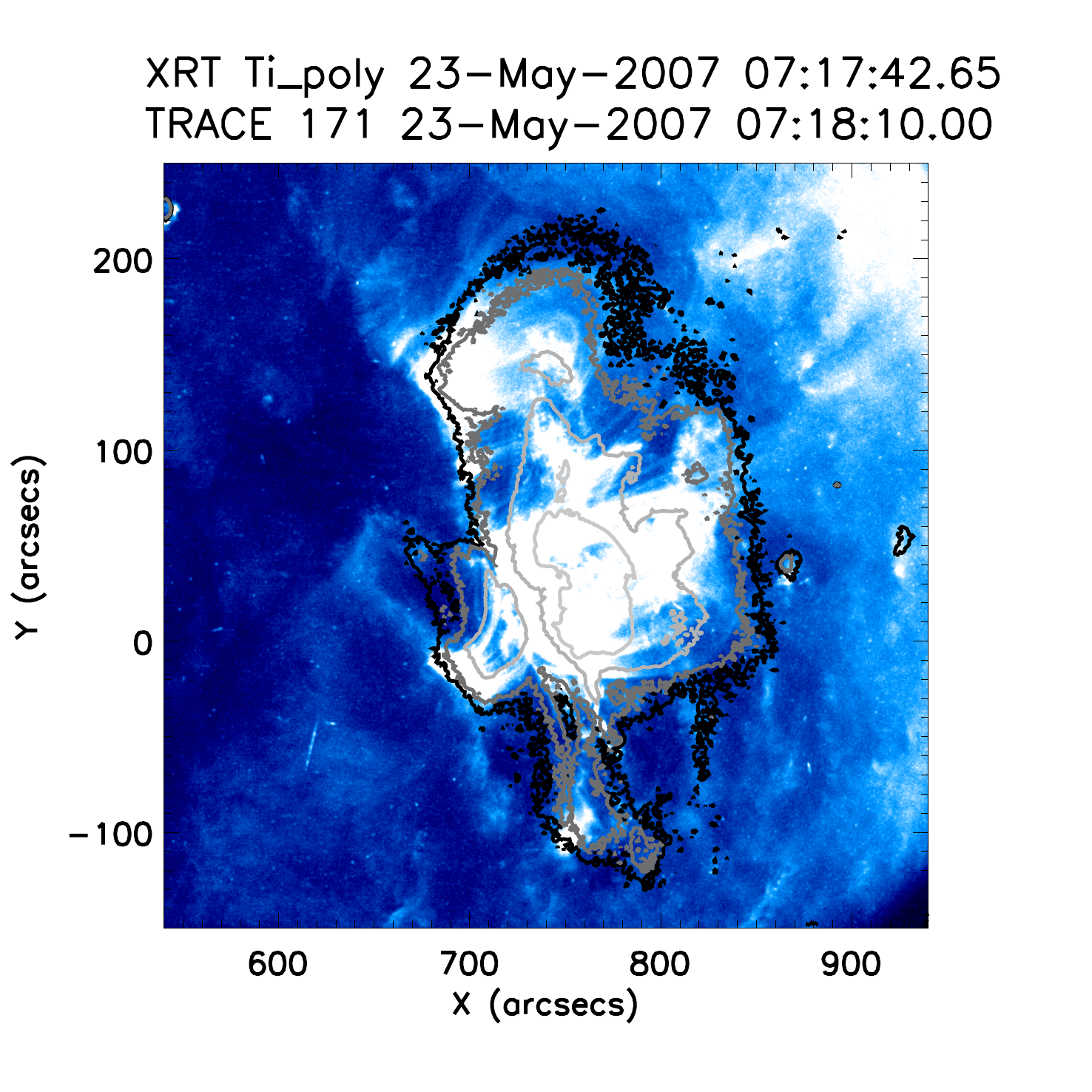} &
\includegraphics[width=70mm]{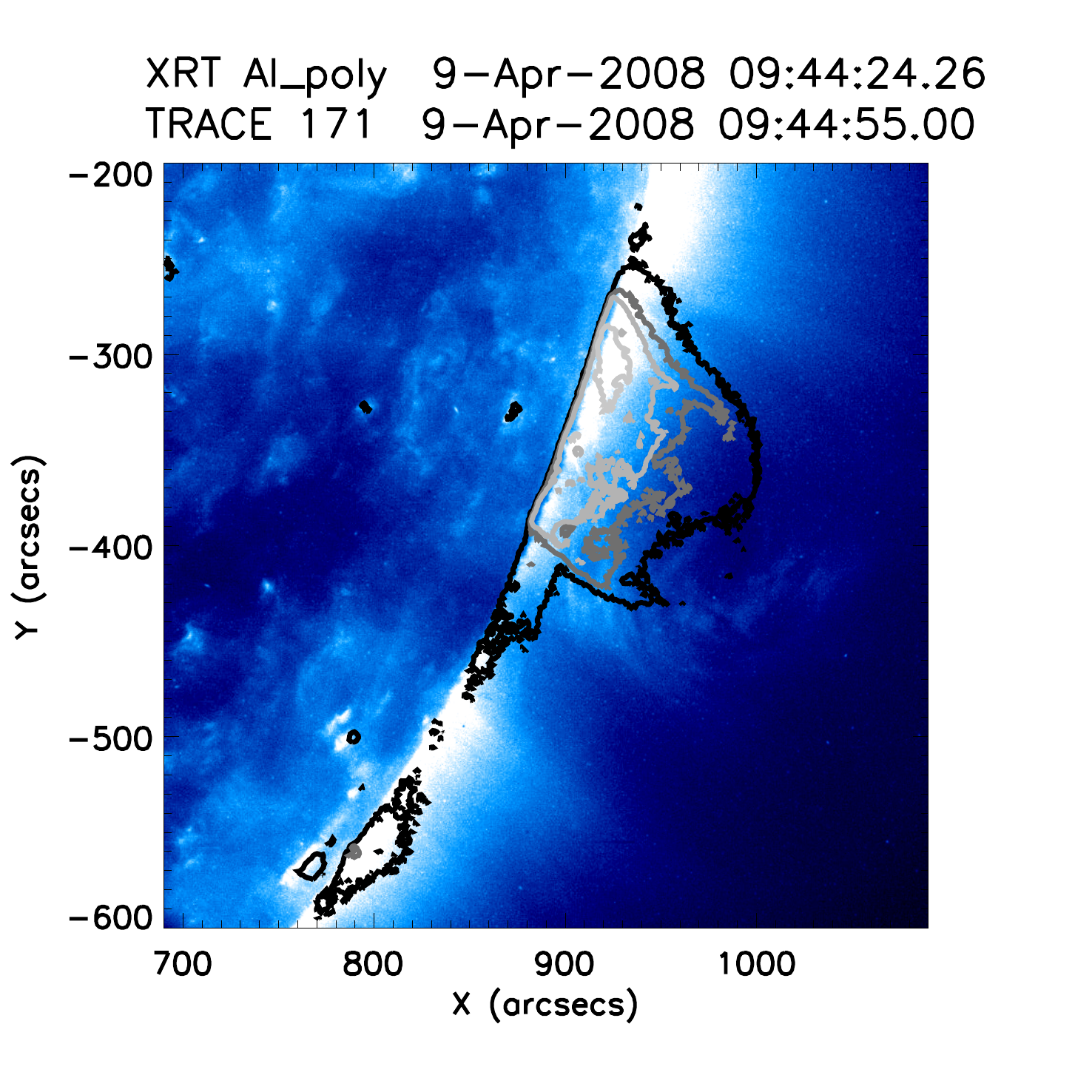}
\end{tabular}
\caption{XRT and TRACE observations on 2007 May 23 (left panels) and 2008 April 09 (right panels) events. 
Arrows indicate the erupting plasmas at the same positions on the XRT and TRACE observations. 
Bottom images show the contours of the XRT observations on the TRACE observations.} 
\label{fig:trace}
\end{figure}

\begin{figure}
\begin{tabular}{c}
\includegraphics[width=140mm]{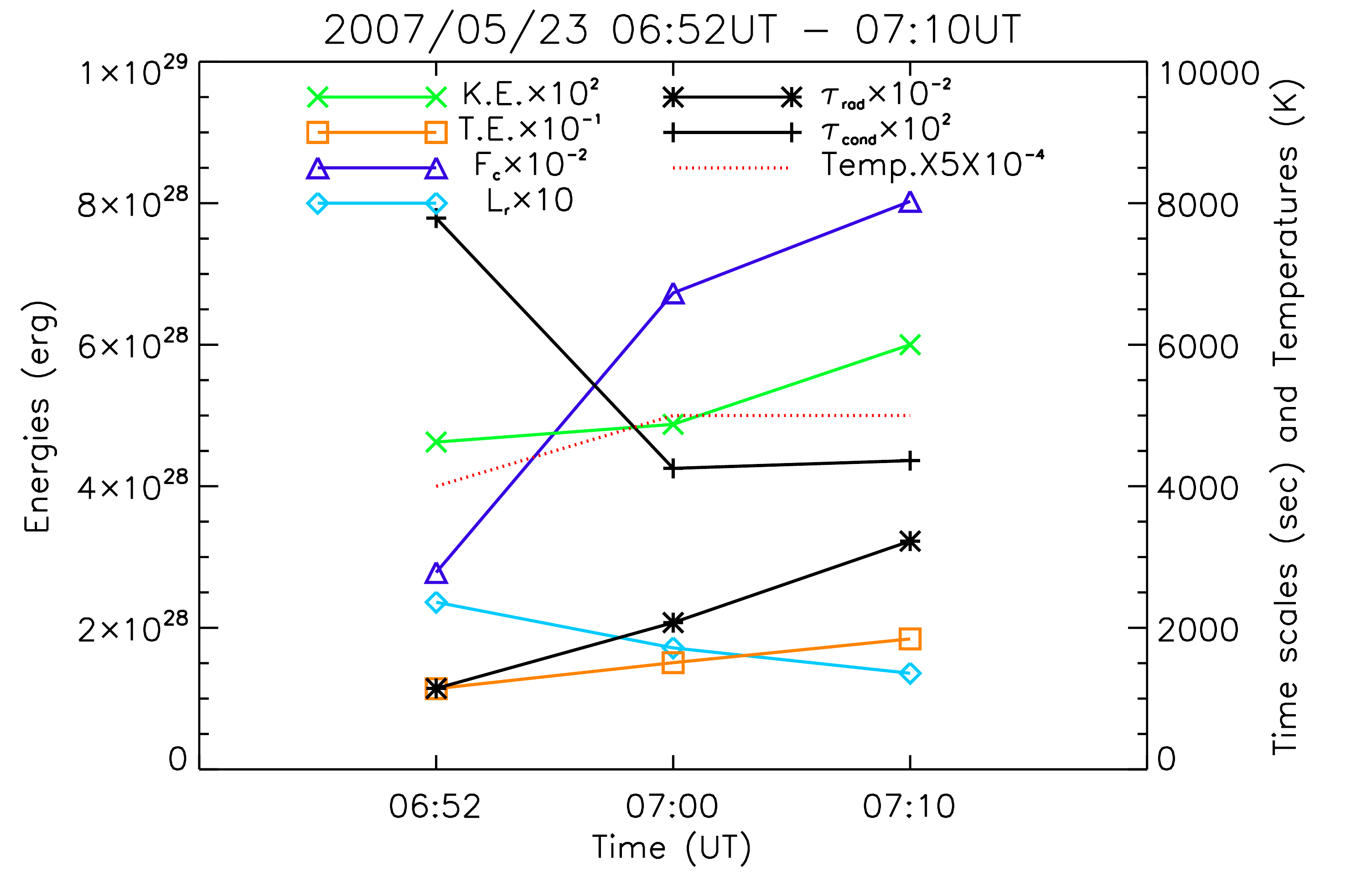} \\
\includegraphics[width=140mm]{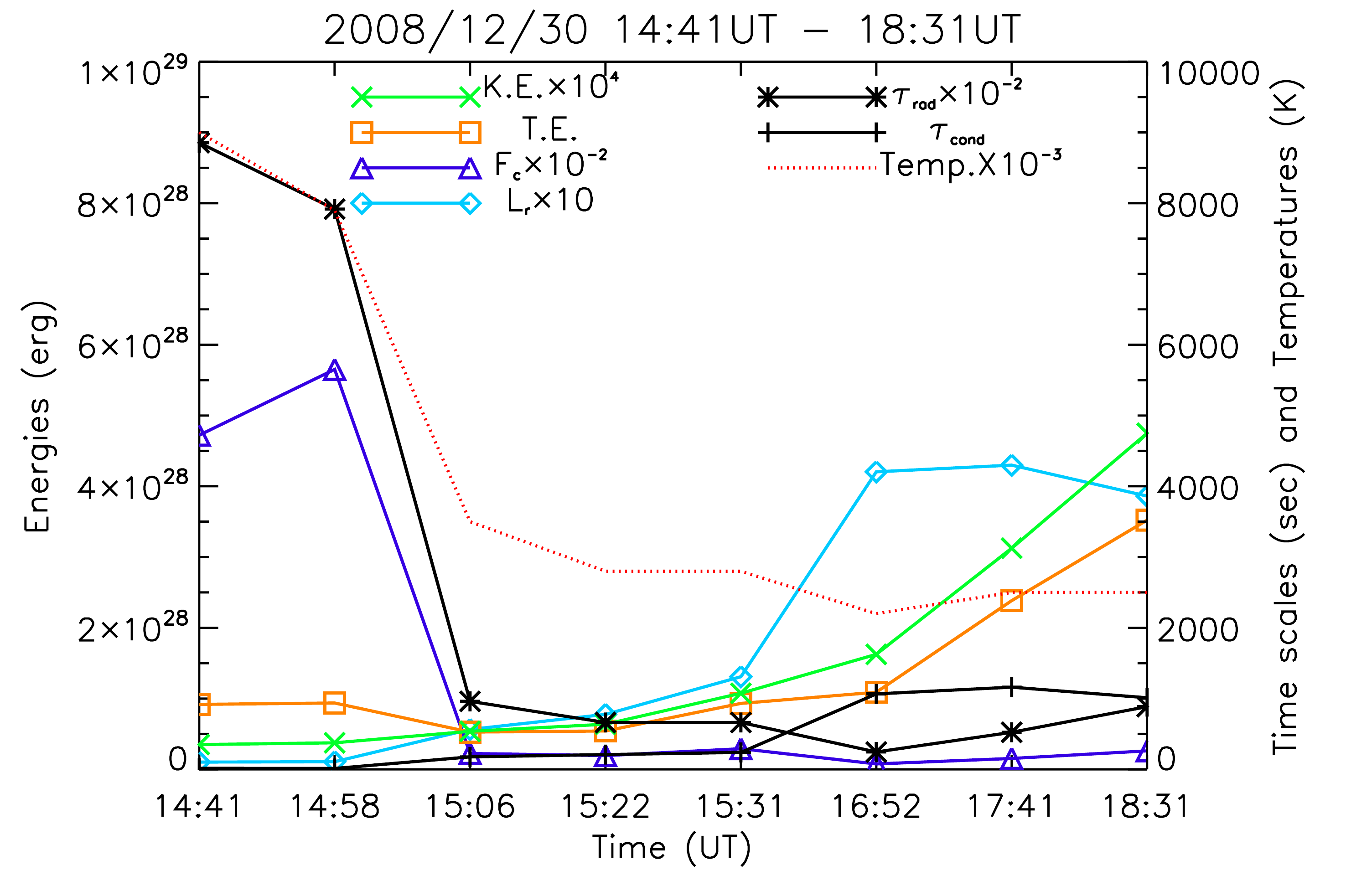}
\end{tabular}
\caption{Energy changes of erupting plasmas on 2007 May 23 and 2008 December 30.} 
\label{fig:energy}
\end{figure}

\end{document}